\documentclass[10pt,twocolumn,letterpaper]{article}

\usepackage[accsupp]{axessibility}  
\usepackage{iccv}
\usepackage{times}
\usepackage{epsfig}
\usepackage{graphicx}
\usepackage{amsmath}
\usepackage{amssymb}
\usepackage{booktabs}
\usepackage{multicol}
\usepackage{multirow}
\usepackage[table]{xcolor}

\usepackage{cite}
\usepackage{algorithmic}
\usepackage{graphicx}
\usepackage{textcomp}
\usepackage[table]{xcolor}
\usepackage{booktabs}
\usepackage[numbers]{natbib}
\usepackage[ruled,vlined]{algorithm2e}
\usepackage{dsfont}
\usepackage{slashed}
\usepackage{arydshln}
\usepackage{float}
\usepackage{pifont}
\usepackage{bbm}
\usepackage{enumitem}
\usepackage{pgfplots}
\usepackage{tikz}

\newcommand{\method}{MPCViT}

\newcommand{\tabincell}[2]{\begin{tabular}{@{}#1@{}}#2\end{tabular}}

\usepackage[pagebackref=true,breaklinks=true,letterpaper=true,colorlinks,bookmarks=false]{hyperref}

\iccvfinalcopy 

\def\iccvPaperID{10727} 

\ificcvfinal\fi

\begin{document}

\title{MPCViT: Searching for Accurate and Efficient MPC-Friendly\\ Vision Transformer with Heterogeneous Attention}



\author {
    Wenxuan Zeng\textsuperscript{\rm 1}\quad
    Meng Li\textsuperscript{\rm 1\thanks{Corresponding author.}}\quad
    Wenjie Xiong\textsuperscript{\rm 2}\quad 
    Tong Tong\textsuperscript{\rm 1}\quad 
    Wen-jie Lu\textsuperscript{\rm 3}\quad \\
    Jin Tan\textsuperscript{\rm 3}\quad
    Runsheng Wang\textsuperscript{\rm 1}\quad
    Ru Huang\textsuperscript{\rm 1}
    \\
    \textsuperscript{\rm 1}Peking University\quad
    \textsuperscript{\rm 2}Virginia Tech\quad
    \textsuperscript{\rm 3}Ant Group\quad
    \\
    \small\texttt{\{zwx.andy,tongtong\}@stu.pku.edu.cn, \{meng.li,wrs,ruhuang\}@pku.edu.cn,}
    \\
    \small\texttt{wenjiex@vt.edu, \{juhou.lwj,tanjin.tj\}@antgroup.com}
}

\maketitle
\ificcvfinal\thispagestyle{empty}\fi

\begin{abstract}
Secure multi-party computation (MPC) enables computation directly on encrypted data and protects both data and model privacy in deep learning inference.
However, existing neural network architectures, including Vision Transformers (ViTs), are not designed or optimized for MPC and incur significant latency overhead. 
We observe Softmax accounts for the major latency bottleneck due to a high communication complexity,
but can be selectively replaced or linearized without compromising the model accuracy.
Hence, in this paper, we propose an MPC-friendly ViT, dubbed \method, to enable accurate yet efficient ViT inference in MPC.
Based on a systematic latency and accuracy evaluation of the Softmax attention and other attention variants,
we propose a heterogeneous attention optimization space.
We also develop a simple yet effective MPC-aware neural architecture search algorithm for fast Pareto optimization.
To further boost the inference efficiency, we propose \method$^+$, to jointly optimize the Softmax attention and other network components, including GeLU, matrix multiplication, etc.
With extensive experiments, we demonstrate that \method~achieves 1.9\%, 1.3\% and 3.6\% higher accuracy with 6.2$\times$, 2.9$\times$ and 1.9$\times$ latency reduction compared with baseline ViT, MPCFormer and THE-X on the Tiny-ImageNet dataset, respectively.
\method$^+$~further achieves a better Pareto front compared with \method.
The code and models for evaluation are available at \url{https://github.com/PKU-SEC-Lab/mpcvit}.

\end{abstract}
\section{Introduction}

\begin{figure}[!tbp]
    \centering
    \includegraphics[width=\linewidth]{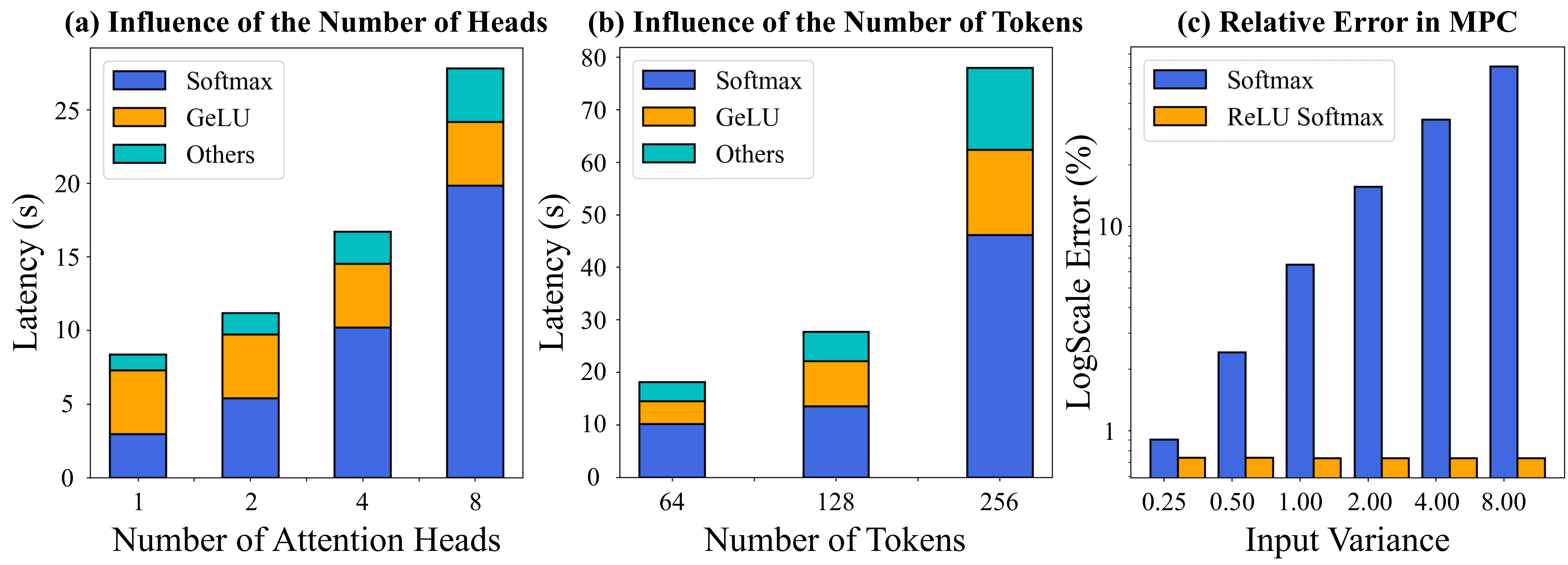}
    \caption{The latency breakdown of a Transformer block w.r.t (a) \# heads and (b) \# tokens; and (c) the relative error of Softmax and ReLU Softmax with different input variances.}
    \label{fig:intro_bar}
\end{figure}

As machine learning models are handling increasingly sensitive data
and tasks, privacy has become one of the major concerns during the model deployment.
Secure multi-party computation (MPC)~\cite{goldreich1998secure} can protect
the privacy of both data and deep neural network (DNN) models and has gained
a lot of attention in recent years
\cite{mishra2020delphi,wang2022characterization,rathee2020cryptflow2}.

\begin{figure*}[!tbp]
    \centering
    \includegraphics[width=0.9\linewidth]{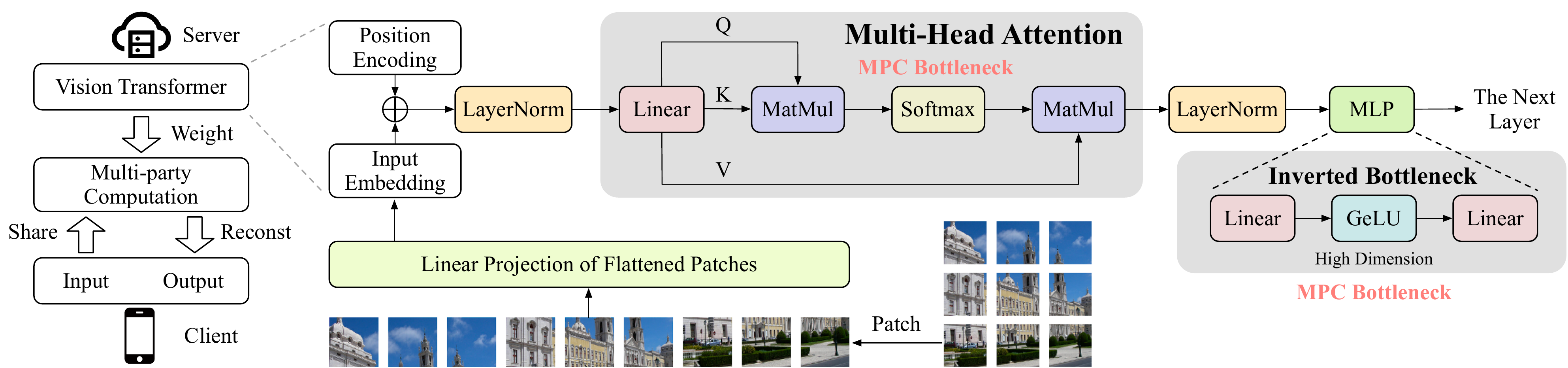}
    \caption{An overall illustration of private Vision Transformer inference in the MPC framework. 
    }
    \label{fig:private_transformer}
\end{figure*}

However, existing DNN architectures, especially the recently proposed Vision Transformers (ViTs) \cite{dosovitskiy2020image,arnab2021vivit,zhang2021multi},
are not designed or optimized for MPC (the high-level private ViT inference framework in MPC is shown in Figure \ref{fig:private_transformer} and Appendix \ref{supp:overview}). 
Although ViTs have achieved superior performance for various vision tasks \cite{gong2021nasvit,zhang2021multi,arnab2021vivit,gu2022multi},
they face several realistic limitations when directly deployed in MPC:
\textbf{\underline{1)} communication overhead:} in contrast to regular inference on plaintext, operations like Softmax, GeLU, max, etc,
require intensive communication in MPC, which usually dominates the total inference latency \cite{mishra2020delphi,wang2022characterization}. 
For example, Softmax is usually very lightweight in plaintext inference.
However, as shown in Figure~\ref{fig:intro_bar}(a) and Figure~\ref{fig:intro_bar}(b), it accounts for the majority of the Transformer inference latency due to the high communication complexity;
\textbf{\underline{2)} approximation error:}  operations like exponential, tanh, reciprocal, etc, cannot be computed directly and require iterative approximation,
limiting the computation accuracy.
For instance, as shown in Figure~\ref{fig:intro_bar}(c), the relative error of Softmax significantly increases when the input variance is large due to its narrow dynamic range \cite{wang2022characterization}.
In contrast, replacing exponential in Softmax with ReLU reduces the relative error drastically (denoted as ReLU Softmax in $\S$\ref{subsec:mpc}).

Numerous efficient Transformer variants have been proposed in recent years
\cite{kitaev2020reformer,wang2020linformer,qin2022cosformer,brown2022dartformer}.
Linear Transformers, including Linformer~\cite{wang2020linformer}, cosFormer~\cite{qin2022cosformer},
Reformer~\cite{kitaev2020reformer}, etc, reduce the quadratic computation complexity of the attentions
and significantly accelerate model inference for long sequences.
However, these Transformer optimizations have primarily focused on reducing the network computation.
Hence, they either retain complex non-linear functions
or still require iterative approximation, both resulting in intensive communication overhead in MPC.
Recently, MPCFormer~\cite{li2022mpcformer} and THE-X~\cite{chen-etal-2022-x}
are proposed to improve the inference efficiency of BERT~\cite{devlin2018bert}
in MPC. MPCFormer simplifies Softmax by replacing exponential with a more MPC-friendly quadratic operation
while THE-X approximates Softmax with a small estimation network.
However, their attention variants is not flexible and can still be too expensive for 
certain latency constraints.
Moreover, the network-level optimization suffers from a large accuracy degradation when directly applied to vision tasks.


In this work, we first breakdown Softmax into more atomic
operations and analyze their impact on the inference accuracy and efficiency,
based on which we propose and compare a set of attention variants comprehensively.
The comparison enables us to find MPC-friendly attention variants, which are either highly accurate or highly efficient.
We further observe not all attentions are equally important within a ViT.
Based on these observations, we propose the first MPC-friendly ViT architecture, dubbed \method.
\method~features a heterogeneous attention design space to explore the trade-off between accuracy and efficiency,
and an MPC-aware differentiable neural architecture search (NAS) algorithm for effective Pareto optimization.
Our contributions can be summarized as follows:
\begin{itemize}
    \item We breakdown Softmax and reveal the impact of atomic operations on the inference accuracy and efficiency,
    based on which a heterogeneous attention design space is proposed.
    \item We propose an effective MPC-aware differentiable NAS algorithm to explore the attention design space based on real latency measurements instead of proxies.
    To the best of our knowledge, we make the first attempt to introduce the MPC-aware NAS to optimize ViT inference in MPC.
    \item Our MPC-friendly ViT model family, \method, outperforms prior-art models in MPC in terms of both accuracy and efficiency.
    \method~achieves 1.9\% and 1.3\% higher accuracy with 6.2$\times$ and 2.9$\times$ latency reduction compared with baseline ViT and MPCFormer on Tiny-ImageNet dataset, respectively.
    \item We further extend \method~to optimize other Transformer components simultaneously with Softmax, named \method$^+$, achieving an even better Pareto front compared with \method.
\end{itemize}

\section{Preliminaries and Related Works}
\label{sec:background}

\subsection{ViT and Efficient Attention}

\textbf{ViT architecture.}
ViT~\cite{dosovitskiy2020image} has demonstrated a great potential to capture the long-range visual dependencies.
It takes image patches as input and is composed of an input projection layer, a stack of Transformer layers, and a task-specific multi-layer
perceptron (MLP) head. 
Each layer consists of a multi-head attention (MHA) layer and an MLP block. The core computation is the Softmax attention:
\begin{equation*}
    \mathrm{Attention}(Q, K, V) = \mathrm{Softmax}(\frac{QK^T}{\sqrt{d_k}})V,
\end{equation*}
where $Q, K, V$ are queries, keys and values, respectively. $d_k$ denotes the embedding dimension of each key, and Softmax normalizes the attention map. 
The detailed explanation of ViT architecture is shown in Figure \ref{fig:private_transformer} and Appendix \ref{supp:vit_arch}. 

\begin{table*}[!tbp]
    \centering
    \caption{Comparison among existing network optimizations for the efficient private inference.}
    \label{tab:comp_related}
    \scalebox{0.8}{
    \begin{tabular}{cccccc}
    \toprule
       Method  & Model & Technique & Optimized Component & Granularity & Linear Fusion \\
       \hline
       Delphi~\cite{mishra2020delphi} & CNN & \tabincell{c}{ReLU-aware NAS} & ReLU & Layer & \textcolor{red}{\ding{55}} \\
       DeepReDuce~\cite{jha2021deepreduce} & CNN & Manually removal & ReLU & Layer & \textcolor{red}{\ding{55}} \\
       CryptoNAS~\cite{ghodsi2020cryptonas} & CNN & ReLU-aware NAS & ReLU & Layer & \textcolor{red}{\ding{55}} \\
       SNL~\cite{cho2022selective} & CNN & ReLU-aware NAS & ReLU & Channel, pixel & \textcolor{red}{\ding{55}} \\
       Sphynx~\cite{cho2021sphynx} & CNN & Block-level NAS & ReLU & Block & \textcolor{red}{\ding{55}} \\
       SENet~\cite{kundu2023learning} & CNN & Sensitivity-aware alloc. & ReLU & Layer, pixel & \textcolor{red}{\ding{55}} \\ 
       RRNet~\cite{peng2023rrnet} & CNN & Hardware-aware NAS & ReLU &  Layer & \textcolor{red}{\ding{55}} \\
       DeepReShape~\cite{jha2023deepreshape} & CNN &  ReLU-equalization & ReLU & Stage & \textcolor{red}{\ding{55}} \\
       SAFENet~\cite{kundu2023learning} & CNN & Mixed-precision poly. approx. & ReLU & Channel  & \textcolor{red}{\ding{55}} \\
       \hdashline
       THE-X~\cite{chen-etal-2022-x} & BERT & Polynomial approx. & GeLU, Softmax, LayerNorm & Network & \textcolor{red}{\ding{55}} \\
       MPCFormer~\cite{li2022mpcformer} & BERT & Polynomial approx. & Softmax, GeLU & Network & \textcolor{red}{\ding{55}} \\
       \hdashline
       \method~(ours)  & ViT & \tabincell{c}{MPC-aware NAS\\Heterogeneous attention} & Softmax & Layer, head, row & \textcolor{red}{\ding{55}} \\
       \hdashline
       \method$^+$ (ours) & ViT & \tabincell{c}{MPC-aware NAS\\Heterogeneous attention\\GeLU linearization\\Linear fusion} & Softmax, GeLU, MatMul & Layer, head, row, token & $\textcolor{green}{\checkmark}$ \\
    \bottomrule
    \end{tabular}
    }
\end{table*}

\textbf{Linear attention.}
Linear attention has been widely studied in previous works \cite{kitaev2020reformer,qin2022cosformer,brown2022dartformer,lu2021soft}.
These works aim at reducing the quadratic increase of compute and memory by leveraging special kernel functions.
Different kernel functions have been proposed in existing works~\cite{song2021ufo,wang2020linformer,qin2022cosformer,bolya2022hydra}.
UFO-ViT~\cite{song2021ufo} proposes to use $\ell2$-Norm as the kernel
function while 
CosFormer~\cite{qin2022cosformer} leverages the cosine distance kernel.
Linformer~\cite{wang2020linformer} approximates self-attention by a low-rank matrix.
Hydra Attention~\cite{bolya2022hydra} proposes to use as many heads
as possible to further reduce the computation complexity of 
the linear attention.
However, these works  primarily
focus on reducing the network computation rather than the communication overhead and are not beneficial for ViT inference in MPC.

\textbf{Non-local neural networks and Scaling attention.}
\cite{wang2018non} presents a generic non-local operation to capture long-range dependencies. The formulation is defined as
$
    \frac{1}{\mathcal{C}(x)}\sum_{\forall j}f(x_i, x_j)g(x_j),
$
where $x$ denotes the input feature map and $f(\cdot)$ is a similarity function, e.g., cosine or Euclidean distance. $g(\cdot)$ computes a certain input representation, and $\mathcal{C}(x)$ is a normalization factor.
By setting $f(\cdot)$ as the dot-product similarity and $\mathcal{C}(x) = n$, where $n$ is the sequence length,
we get an attention variant, named Scaling Attention (ScaleAttn).
ScaleAttn is an extremely simple attention 
and only involves linear operations like multiplication and addition. ScaleAttn can be further re-parameterized to accelerate computation\footnote{In this work, we do not focus on computational overhead in MPC since communication dominates the overhead.} as below:
\begin{equation*}
  \mathrm{ScaleAttn}(Q, K, V) = \frac{1}{n}(QK^T)V 
  = \frac{Q}{\sqrt{n}}(\frac{K^T}{\sqrt{n}}V). 
\end{equation*}

\subsection{Multi-Party Computation}
\label{subsec:mpc}

\begin{table*}[!tbp]
    \center
    \caption{Top-1 accuracy and latency comparison of different attention variants on CIFAR-10. The rows in grey are attention variants with the lowest latency and the highest accuracy, respectively. We also show the properties and remaining operations of each variant.}
    \label{tab:attention_variants_operation}
    \scalebox{0.75}{
        \begin{tabular}{c|ccc|ccc|cc}
        \toprule
        Attention Variant & Monotonicity & Non-negativity & Normalization & Exponential & Max & Reciprocal & Acc. (\%)  & Lat. (s) \\
        \hline
        Softmax Attention \cite{dosovitskiy2020image}     & \textcolor{green}{\checkmark} & \textcolor{green}{\checkmark} & \textcolor{green}{\checkmark} & \textcolor{green}{\checkmark} & \textcolor{green}{\checkmark} & \textcolor{green}{\checkmark}   & 92.69 & 6.82 \\
        \hdashline
        Linformer Attention~\cite{wang2020linformer} & \textcolor{green}{\checkmark} & \textcolor{green}{\checkmark} & \textcolor{green}{\checkmark} & \textcolor{green}{\checkmark} & \textcolor{green}{\checkmark} & \textcolor{green}{\checkmark} & 90.85 & 5.44
        \\
        \hdashline
        ReLU Attention        & \textcolor{green}{\checkmark} & \textcolor{green}{\checkmark} & \textcolor{red}{\ding{55}}  & \textcolor{red}{\ding{55}} & \textcolor{green}{\checkmark} & \textcolor{red}{\ding{55}}  & \texttt{fail}  & \texttt{fail} \\
        ReLU6 Attention       & \textcolor{green}{\checkmark} & \textcolor{green}{\checkmark} & \textcolor{red}{\ding{55}}  & \textcolor{red}{\ding{55}}  & \textcolor{green}{\checkmark} &  \textcolor{red}{\ding{55}}  & 90.50 & 3.02  \\
        Sparsemax Attention~\cite{martins2016softmax} & \textcolor{green}{\checkmark} & \textcolor{green}{\checkmark} & \textcolor{red}{\ding{55}}  & \textcolor{red}{\ding{55}} & \textcolor{green}{\checkmark} & \textcolor{red}{\ding{55}}   & 91.23 & 3.23 \\
        XNorm Attention~\cite{song2021ufo} & \textcolor{green}{\checkmark} & \textcolor{red}{\ding{55}} & \textcolor{red}{\ding{55}} & \textcolor{red}{\ding{55}} &  \textcolor{red}{\ding{55}}  &  \textcolor{green}{\checkmark}    & 91.24 & 13.25  \\
        Square Attention      & \textcolor{red}{\ding{55}} & \textcolor{green}{\checkmark} & \textcolor{red}{\ding{55}} & \textcolor{red}{\ding{55}}  & \textcolor{red}{\ding{55}}  &  \textcolor{red}{\ding{55}}  & 91.27 & 0.72 \\
        2Quad Attention~\cite{li2022mpcformer} & \textcolor{red}{\ding{55}} & \textcolor{green}{\checkmark} & \textcolor{green}{\checkmark} & \textcolor{red}{\ding{55}}  & \textcolor{red}{\ding{55}}  & \textcolor{green}{\checkmark}   &  91.86 & 4.22 \\
        \hdashline
        \rowcolor{gray!20}
        Scaling Attention (ScaleAttn)~\cite{wang2018non} & \textcolor{green}{\checkmark} & \textcolor{red}{\ding{55}} & \textcolor{red}{\ding{55}} & \textcolor{red}{\ding{55}} & \textcolor{red}{\ding{55}} & \textcolor{red}{\ding{55}}  &   91.52 & \textbf{0.66} \\
        \rowcolor{gray!20}
        ReLU Softmax Attention (RSAttn)~\cite{mohassel2017secureml} & \textcolor{green}{\checkmark} & \textcolor{green}{\checkmark} & \textcolor{green}{\checkmark} & \textcolor{red}{\ding{55}} & \textcolor{green}{\checkmark}  & \textcolor{green}{\checkmark}  &   \textbf{92.31} & 5.32 \\
        \bottomrule
        \end{tabular}
    }
\end{table*}

MPC~\cite{goldreich1998secure} is a cryptographic technique 
that offers a promising solution to protect the privacy of both data and model during deep learning inference.
Appendix \ref{supp:crypto} depicts the cryptographic primitives for MPC, including Secret Sharing, Oblivious Transfer and Garble Circuit.
~\cite{goldreich1998secure} proposes an efficient MPC protocol for addition and multiplication.
\cite{mohassel2017secureml} proposes efficient two-party computation
protocols for various arithmetic operations and approximations for
non-linear functions in DNN. For example, \cite{mohassel2017secureml} approximates Softmax with ReLU Softmax, i.e.,
\begin{align*}
    \mathrm{ReLUSoftmax}(x) = \frac{\mathrm{ReLU}(x_i)}{\sum_i \mathrm{ReLU}(x_i) + \epsilon}, 
\end{align*}
where $\epsilon$ is a very small value to avoid the zero denominator.
ABY3~\cite{mohassel2018aby3} is proposed to enable efficient conversion between different secret sharing schemes in three-party computation.
Delphi~\cite{mishra2020delphi} develops a hybrid MPC protocol and optimizes the network to explore the performance-accuracy trade-off.
Iron \cite{meng2022iron} proposes a set of protocols to reduce the communication
overhead of matrix-matrix multiplications (MatMuls), LayerNorm, etc for Transformer-based models.

Although MPC protocols have been improved significantly in recent
years, DNN inference in MPC still suffers from intensive communication and latency overhead compared to the plaintext inference.
To reduce the latency, different algorithms have been proposed 
to optimize convolutional NNs (CNNs) with a focus on ReLU.
We compare the important characteristics of our methods with existing related work in Table \ref{tab:comp_related}.
SNL~\cite{cho2022selective} and DeepReDuce~\cite{jha2021deepreduce} propose to
remove ReLUs and linearize a subset of neurons selectively.
SAFENet~\cite{lou2020safenet}, Delphi~\cite{mishra2020delphi}, RRNet~\cite{peng2023rrnet} and PolyMPCNet \cite{peng2022polympcnet} replace ReLUs with polynomial functions that are more MPC-friendly via NAS for CNNs.
Sphynx \cite{cho2021sphynx} further develops an MPC-friendly architecture
search space for CNNs and leverages NAS to optimize the accuracy
given different latency constraints.
However, these works mainly focus on the ReLU counts in CNNs,
and accelerating ViTs with latency consideration is still a challenge.
MPCFormer \cite{li2022mpcformer} and THE-X \cite{chen-etal-2022-x} recently propose to replace Softmax with more MPC-friendly operations for BERT, but the network-level method incurs 
limited latency reduction and a large accuracy degradation.
As shown in Table \ref{tab:comp_related}, our work proposes an MPC-aware NAS to explore a fined-grained heterogeneous attention optimization space, and further presents a GeLU linearization technique to reduce the latency of MLP.
\section{Motivating Inspiration of \method}
\label{sec:motivating}


In this section, we breakdown Softmax for accuracy and efficiency evaluation
and compare two ways of building MPC-friendly ViTs.
The analysis serves as our motivation of proposing \method,
which features heterogeneous attention and NAS-based Pareto optimization.

\begin{figure}[!tbp]
    \centering
    \includegraphics[width=\linewidth]{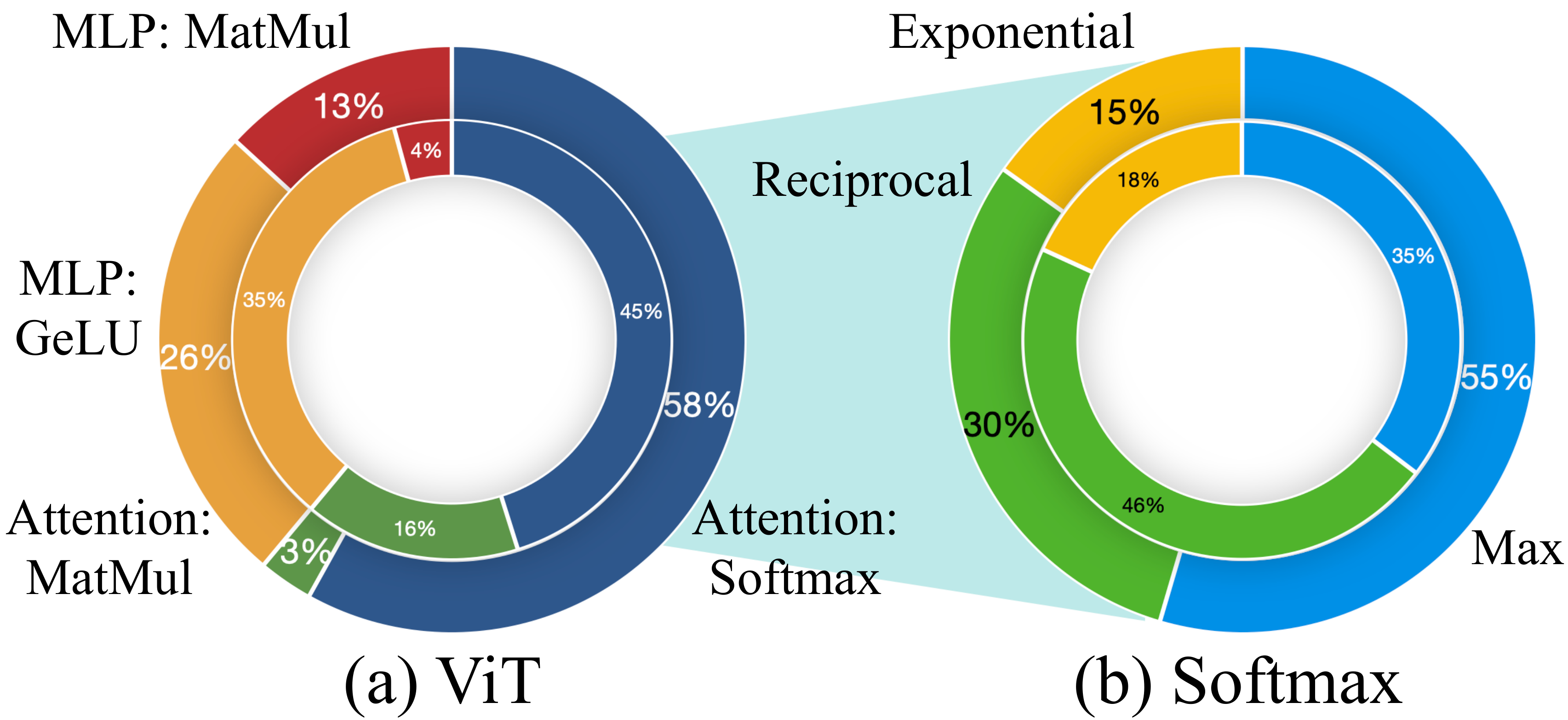}
    \caption{(a) Latency breakdown of a 4-head ViT with 256 hidden dimension and 65 tokens on CIFAR-10 dataset; (b) latency breakdown of Softmax. Both (a) and (b) consist of SEMI-2K (outside) and Cheetah (inside) protocols.}
    \label{fig:latency_breakdown}
\end{figure}

\textbf{Motivation 1: ViT latency bottleneck in MPC.}
In Figure~\ref{fig:intro_bar}, we profile the latency of a Transformer block
with different number of heads and tokens. We now further dive into 
Softmax and evaluate its latency breakdown. 
We profile a 4-head ViT with 65 tokens and 256 hidden dimension
assuming the WAN setting as defined in $\S$\ref{sec:setup}. We consider two
widely used protocols, including SEMI-2K \cite{cramer2018spdz2k} and Cheetah~\cite{huang2022cheetah}.
From both Figure~\ref{fig:intro_bar} and Figure~\ref{fig:latency_breakdown}, we make the following
observations:
    \underline{1)} Softmax is the major latency bottleneck in most cases.
    The latency of GeLU and MatMuls also becomes non-negligible when the number of heads or tokens is small;
    \underline{2)} Softmax mainly involves three operations, including exponential,
    reciprocal and max. All of the operations are important for the latency overhead.
Therefore,
we are motivated to get rid of these expensive operations as many as possible, and
we propose to first focus on optimizing
the Softmax and only consider GeLU and MatMul when the number of heads or tokens is small.

\textbf{Motivation 2: comparison of different attention variants.}
To reduce the latency of Softmax in the attention module,
there are two different ways, i.e., \underline{1)} reducing the dimension of
$QK^T$ such as Linformer~\cite{wang2020linformer},
which approximates $QK^T$ with a low rank matrix;
and \underline{2)} simplifying Softmax by replacing exponential, reciprocal or max with other operations.
For the second method, 
while the possible activation variants can be abundant,
we observe the following three important properties of Softmax, which helps us
to select promising candidates: \textit{monotonicity, non-negativity, and
normalized sum to 1}. 
The first two properties are related to the exponential while the last one is achieved with reciprocal.
We propose and evaluate the following variants, including 
ReLU, ReLU6, Sparsemax~\cite{martins2016softmax}, XNorm~\cite{song2021ufo},
Square, Scaling~\cite{wang2018non}, 2Quard~\cite{li2022mpcformer}, Linformer~\cite{wang2020linformer} and ReLU Softmax~\cite{mohassel2017secureml}. 
We define the attention with ReLU Softmax as RSAttn for short.
The formulation of these attention variants is described in Appendix \ref{supp:detail_attn}.

To compare these attention variants, we train each model for 300 epochs on CIFAR-10 and compare the latency, accuracy, remaining operations, and
how they satisfy the properties of Softmax in Table~\ref{tab:attention_variants_operation}.
We make the following observations:
    \underline{1)} Attention with more expensive operations can be more inefficient. For instance, 2Quad has an extra reciprocal compared with Square such that 2Quad suffers from a higher latency;
    \underline{2)} Linformer suffers from a high accuracy degradation ($\sim$1.8\% compared with Softmax attention) as it significantly reduces the matrix dimensions such that the model learning capability is degraded, and the computation reduction of this optimization cannot benefit latency reduction a lot.
    Hence, the first method is not preferred;
    \underline{3)} ReLU Softmax and 2Quad achieve higher accuracy among other variants and
    we hypothesize this is because they both normalize the sum to 1, which is realized
    with reciprocal and also leads to relatively higher latency.
    ReLU Softmax is more preferred over Softmax and 2Quad due to its small approximation error and small accuracy degradation;
    \underline{4)} Scaling is equivalent to directly removing the Softmax. Although it is the fastest, directly linearizing all the Softmax leads to a large accuracy degradation;
    \underline{5)} it is interesting to notice that although XNorm attention has linear computation, it incurs the highest latency because computing $\ell2$-norm involves both reciprocal and square root, both of which are expensive in MPC.

\textbf{Motivation 3: Not all attentions are equally important.}
As observed in \cite{brown2022dartformer,liu2022neural,michel2019sixteen,su2022vitas,voita2019analyzing,pan2022less}, not all attentions are equally important and different layer prefers different attention type.
For instance, \cite{pan2022less} identifies the minor contribution of the early attention layers in recent ViTs, while \cite{li2022q} demonstrates the second head shows higher importance than other heads.
Thus, the importance of attentions among different layers
or even the attentions within the same layer can be different.
It indicates we can safely replace the expensive Softmax or ReLU Softmax with the cheaper Scaling without compromising the network accuracy.

\textbf{Remark} 
The analysis above provide us with the following intuitions when
designing an MPC-friendly ViT: 

\begin{itemize}
    \item[1)] ReLU Softmax is
more preferred over Softmax and more MPC-friendly, and Scaling has the highest inference efficiency;
    \item[2)] by selectively replacing
the ReLU Softmax with the cheaper Scaling, it is possible to reduce the inference latency without compromising the inference accuracy;
    \item[3)] when the latency of GeLU and Softmax are comparable, we have an opportunity to further simplify GeLU along with Softmax.
\end{itemize}

\section{The Proposed \method~Algorithm}

In this section, we introduce \method, which for the first time optimizes the ViT architecture with NAS-based algorithm for a fast private inference in MPC.

\subsection{Overview of Optimization Flow}

\begin{figure}[!tb]
    \centering
    \includegraphics[width=\linewidth]{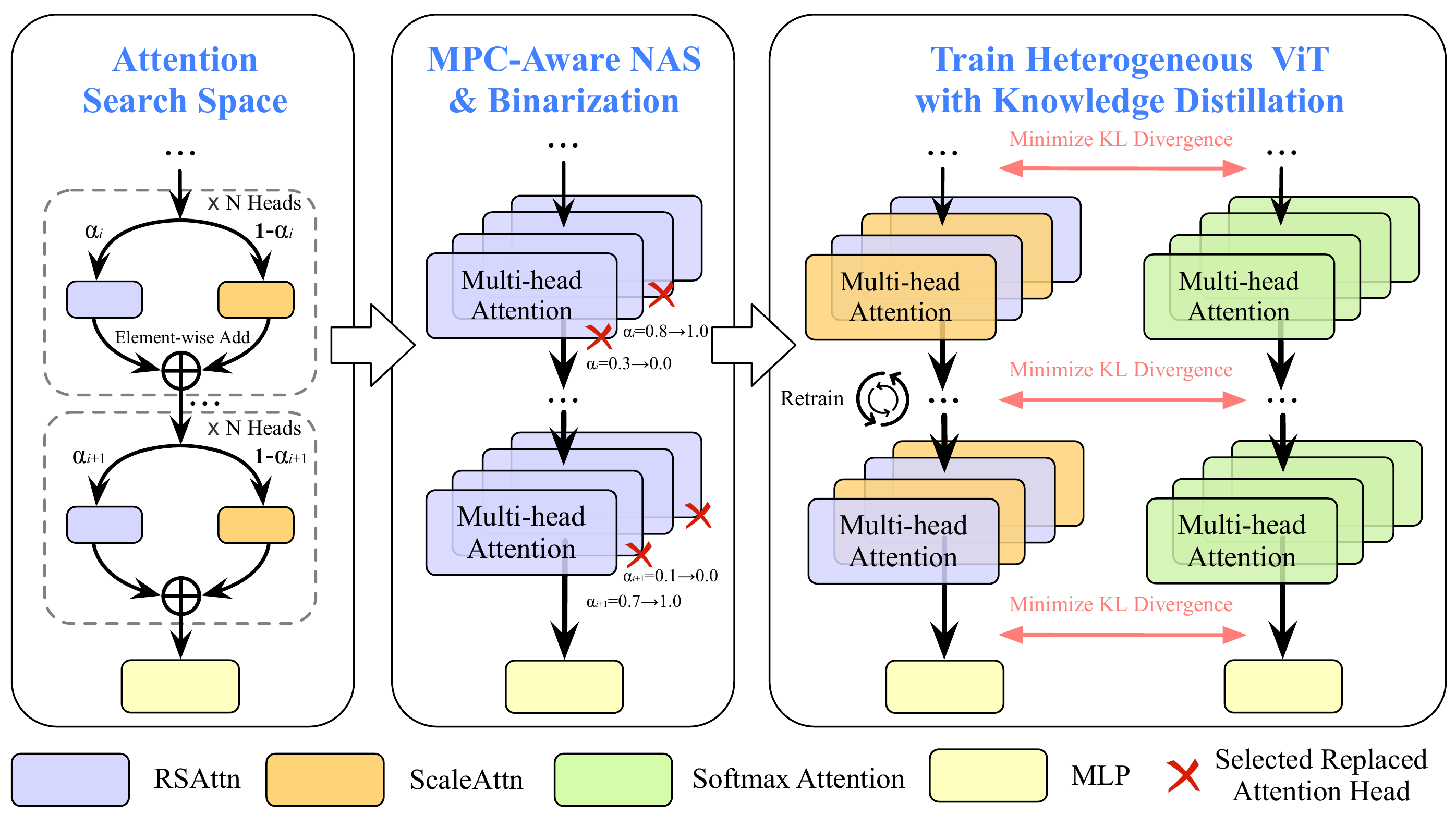}
    \caption{Overview of our proposed \method~pipeline.}
    \label{fig:overview}
\end{figure}

Based on the motivations described above, we now describe our
MPC-aware ViT optimization flow as shown in Figure~\ref{fig:overview}.
We propose to replace the Softmax with the accurate yet MPC-friendly ReLU Softmax in the ViT first and then,
linearize the relatively ``unimportant'' ReLU Softmax with efficient Scaling.
To find the ``unimportant'' ReLU Softmax,
we first design a heterogeneous search space that includes both ReLU Softmax and Scaling attentions.
Architecture parameters are defined to measure the importance of ReLU Softmax and 
enable the selection between the 
two attention variants with different structure granularities 
($\S$\ref{subsec:search_space}).
Then, we propose an MPC-aware differentiable NAS algorithm to learn the architecture
parameters and model parameters simultaneously.
We select between the attention variants based on the architecture parameters
and the latency constraints ($\S$\ref{subsec:nas}).
In the last step, we retrain the ViT with heterogeneous attentions from
scratch. 
We leverage knowledge distillation (KD) to improve the accuracy of
the searched networks ($\S$\ref{subsec:kd}).
Appendix \ref{supp:algo} depicts the complete algorithm flow of \method.

\subsection{Heterogeneous Attention Search Space}
\label{subsec:search_space}

\begin{figure}[!tbp]
    \centering
    \includegraphics[width=\linewidth]{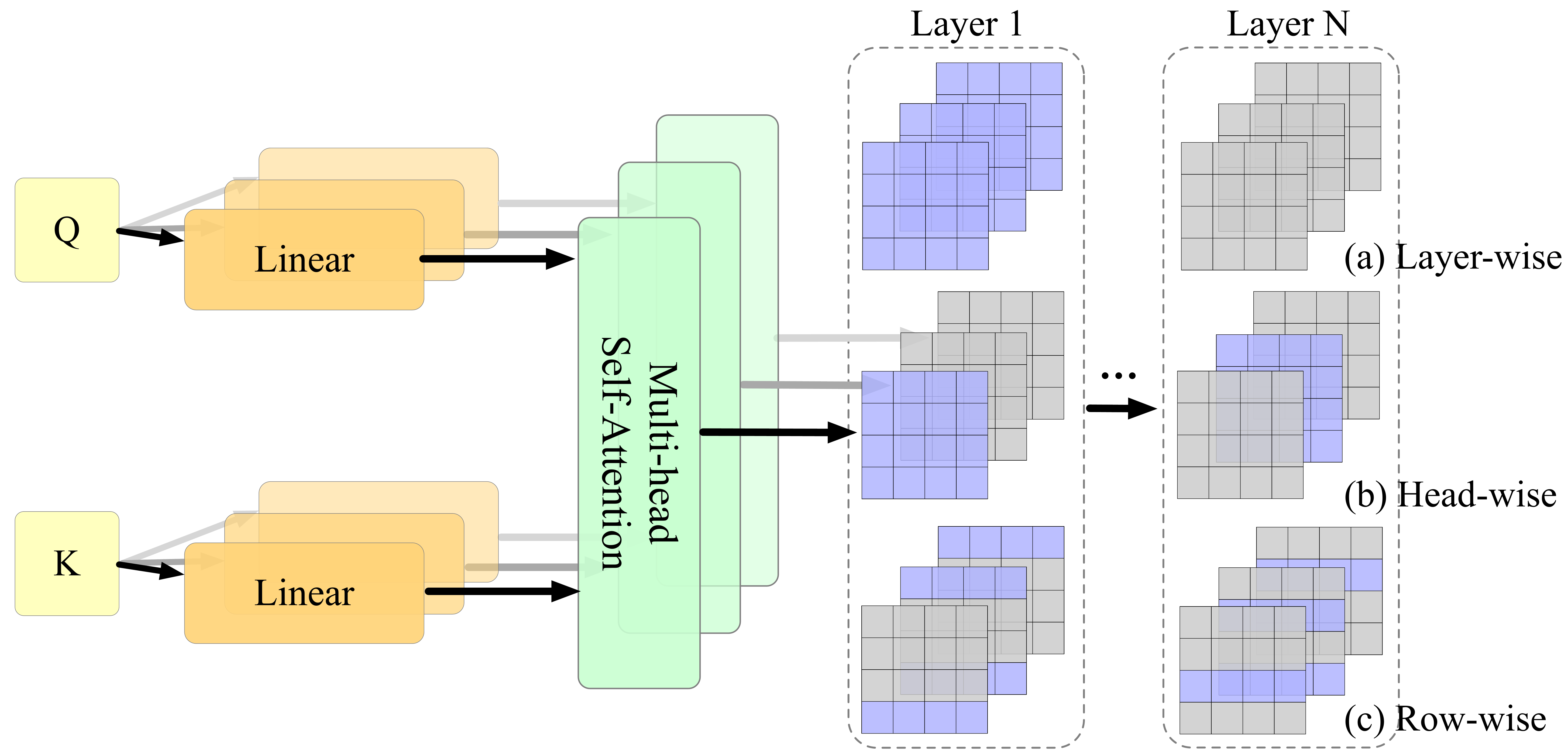}
    \caption{An illustration of our proposed search space with different structure granularities: (a) layer-wise search space with the size of $2^{\mathrm{\# layers}}$; (b) head-wise search space with the size of $2^{\mathrm{\# layers}\times\mathrm{\# heads}}$ and; (c) row-wise search space with the size of $2^{\mathrm{\# layers}\times\mathrm{\# heads}\times\mathrm{\# tokens}}$. The blocks in blue represent RSAttn and the blocks in grey represent ScaleAttn.}
    \label{fig:search_space}
\end{figure}

Our heterogeneous search space combines two candidate attention variants, i.e., MPC-efficient ScaleAttn and
high-accuracy RSAttn and has the following three different structural
granularities as shown in Figure~\ref{fig:search_space}:
\begin{itemize}
    \item \textit{Layer-wise} search space is coarse-grained, 
        with each layer using either ScaleAttn or RSAttn.
        The total number of possible
        architectures in the search space can be limited, especially for
        shallow ViT models.
    \item \textit{Head-wise} search space selects ScaleAttn or RSAttn for each attention head of each layer.
    \item \textit{Row-wise} search space (i.e., token-wise) is the most fine-grained and mixes the two
        attention variants along each row of the attention map.
\end{itemize}

Though the search spaces have different granularities, they share very similar NAS
formulation. In the rest of the explanation, we will use the head-wise search space as
an example to explain our algorithm. 

\subsection{MPC-Aware Differentiable NAS}
\label{subsec:nas}

Many research efforts have been made to NAS including reinforcement learning method (RL) \cite{jaafra2019reinforcement,guo2019irlas,wang2019haq}, however, RL-based NAS requires significant overhead during searching.
Recent years, \cite{wu2019fbnet,jiang2020hardware,tan2019mnasnet,dai2019chamnet} explore hardware-aware NAS algorithms to make model fit the rigorous hardware requirement. 
For private ViT inference, we also need a NAS algorithm that takes MPC conditions into consideration. 

Given the heterogeneous attention search space, we now introduce our simple yet effective MPC-aware differentiable NAS algorithm.

\textbf{Search formulation}. 
Inspired by \cite{brown2022dartformer}, we introduce an architecture parameter
$\alpha$ ($0 \leq \alpha \leq 1$) for each head, which is an auxiliary learnable
variable that helps to choose between RSAttn and ScaleAttn. 
Then, each head
implements the computation below:
\begin{equation*}
    \alpha \cdot \mathrm{ReLUSoftmax}(\frac{QK^T}{\sqrt{d_k}}) V + (1 - \alpha) \cdot \frac{\mathrm{ScaleAttn}(Q, K, V)}{\sqrt{d_k}}.
\end{equation*}

Note that we also add $\frac{1}{\sqrt{d_k}}$ for ScaleAttn to make the search process more stable and robust.

\textbf{Search objective function.} 
We aim at training ViTs with high accuracy but as few RSAttn as possible. 
We introduce
$\ell1$-penalty into the objective function 
and formulate the NAS as a one-level optimization problem. 
To take MPC overhead into consideration, we incorporate realistic inference latency constraints into objective function to construct the MPC-aware NAS algorithm as
\begin{gather*}
    \mathrm{Cost}(\alpha_{ij}) = \alpha_{ij} \cdot \mathrm{Lat}(\mathrm{ATTN}_{ij}), \\
    \min\limits_{\theta, \alpha} \ell(f_{\theta, \alpha}(x), y) + \lambda \cdot (\sum_{i=1}^L \sum_{j=1}^N \lVert \mathrm{Cost}(\alpha_{ij}) \lVert_1),
\end{gather*}
where $x$ and $y$ are the input-label pairs, $\ell(\cdot)$ is the loss function,
$\lambda$ is a hyper-parameter to control the weight of inference overhead, $N$ is the number of heads in one layer, and $L$ is the number of layers. $\lVert \cdot \lVert_1$ means $\ell1$-norm, and $\mathrm{ATTN}_{ij}$ means the candidate attention of $i$-th layer and $j$-th head.
We initialize $\alpha$ for
each head to 1.0 and jointly optimize
the network parameter $\theta$ and architecture parameter 
$\alpha$ during searching. Note that our differentiable algorithm
computes the gradients of $\alpha$ for all the attention heads across
different layers in ViT simultaneously.

\textbf{Architecture parameter binarization}. 
After the network searching, we obtain $\alpha$ for each attention head in 
all layers. To select either RSAttn or ScaleAttn for each head,
we binarize $\alpha$ based on the following rule:
we use the \textit{top-k} rule according to the ratio of RSAttn budeget $\mu$, which is defined as $\mathrm{\# RSAttn}/\mathrm{\# Heads}$.
Specifically, we first find the $\mu LN$-th largest $\alpha$, 
denoted as  $\alpha^*$, and then we binarize $\alpha$ for each
attention head following $\mathbbm{1}(\alpha \geq \alpha^*)$. 
Our proposed search method does not need to train candidate networks, but only needs to train once and then select, which accelerates our search process.
Now, we obtain a heterogeneous ViT,  
and by changing $\mu$, we can obtain a family of ViTs with different
accuracy and efficiency trade-off.

\subsection{Train the Searched ViT for Better Performance}
\label{subsec:kd}

After binarization, RSAttn with a small $\alpha$ are
replaced with ScaleAttn.
We find that directly training a heterogeneous ViT results in an accuracy drop.
So, \textit{how to effectively train the heterogeneous ViT and restore 
the model accuracy} is an important question for us.

\textbf{Token-wise feature-based KD.}
KD~\cite{hinton2015distilling} is a
powerful tool for a student network to learn from a teacher network,
which is usually larger and more complex with a higher learning capacity.
To improve the accuracy of \method,
we leverage token-wise feature-based self-distillation along with a logits-based KD, which uses the baseline ViT as the teacher network while other larger networks
can be easily plugged in and used in our framework.
For the feature-based KD, we take the output features of the last ViT layer
from both the student and the teacher networks and compute the $\ell2$-distance, denoted as $\mathcal{L}_{feature}$.
We also use KL-divergence loss $\mathcal{L}_{KL}$ for the logits-based KD, denoted as $\mathcal{L}_{logits}$.
Combining distillation loss with the task-specific cross-entropy (CE) loss $\mathcal{L}_{CE}$, we obtain the final training objective under a distillation temperature $\mathcal{T}$ as follows:
\begin{equation*}
    \begin{split}
        \mathcal{L}_{train} &= \chi \mathcal{L}_{CE} + \beta \mathcal{L}_{logits} + \gamma \mathcal{L}_{feature} \\
        &= \chi \mathcal{L}_{CE} + \beta \mathcal{L}_{KL}(\sigma(\frac{z_{s}}{\mathcal{T}}), \sigma(\frac{z_{t}}{\mathcal{T}})) + \gamma \lVert z_s-z_t \lVert_2,
    \end{split}
\end{equation*}
where $\chi$, $\beta$ and $\gamma$ are training hyper-parameters to balance different loss functions in the training objective,
$z_s$ and $z_t$ are feature maps from student and teacher,
and $\sigma$ is Softmax.
This self-distillation requires no extra model and only introduces negligible training cost, and has no influence on inference latency or communication overhead in MPC.

\subsection{\method$^+$: Support for GeLU Linearization}
\label{sec:mpcvit+}

\begin{figure}[!tbp]
    \centering
    \includegraphics[width=\linewidth]{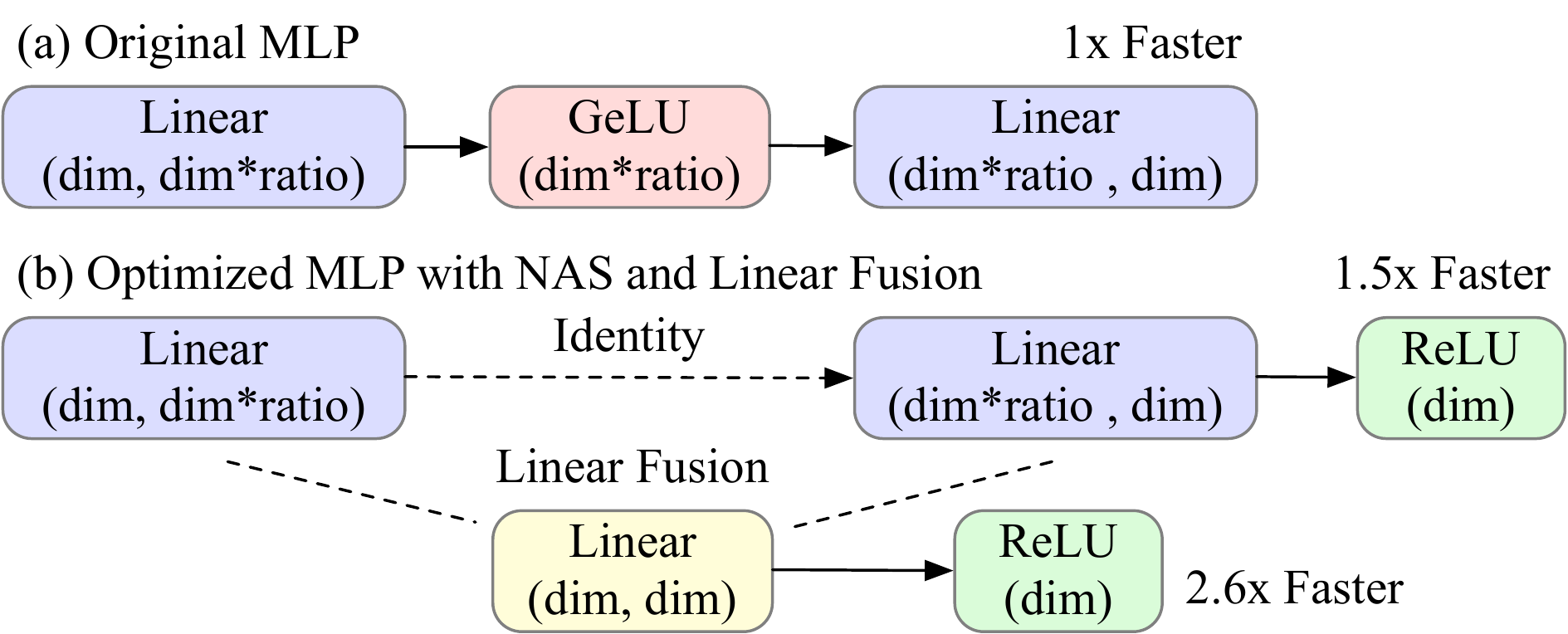}
    \caption{\method$^+$: The optimization of our proposed GeLU linearization and linear fusion for the MLP block.}
    \label{fig:gelu_opt}
\end{figure}

The above design of \method~leverages NAS algorithm to build the efficient yet effective attention mechanism. 
Besides attention, in MLP blocks, non-linear GeLU also impacts the communication cost in some protocols, e.g., Cheetah when the number of heads or tokens is small. And two linear layers near GeLU are also costly because the inverted-bottleneck structure of MLP leads to high-dimensional MatMul, which is shown in Figure \ref{fig:intro_bar} and Figure \ref{fig:latency_breakdown}(a).
To further improve the efficiency of ViTs, we propose an extended version called \method$^+$.
Specifically, \method$^+$ searches and removes relatively ``unimportant'' GeLUs.
Given architecture parameter $\beta$ for GeLU, the NAS algorithm is similar to the attention optimization:
\begin{align*}
    (\beta \cdot \mathrm{GeLU}(x) + (1 - \beta) \cdot x) \cdot \mathcal{W}.
\end{align*}

The architecture parameters $\alpha$ and $\beta$ are jointly learnt during the search, thus the NAS loss is defined along with the classification loss $\min\limits_{\theta, \alpha, \beta} \ell(f_{\theta, \alpha, \beta}(x), y)$:
\begin{gather*}
    \mathrm{Cost}(\beta_{ij}) = \beta_{ij} \cdot \mathrm{Lat}(\mathrm{Act}_{ij}), \\
     \lambda \cdot (\sum_{i=1}^L \sum_{j=1}^N \lVert \mathrm{Cost}(\alpha_{ij}) \lVert_1)
    + \eta \cdot (\sum_{i=1}^L \sum_{j=1}^T \lVert \mathrm{Cost}(\beta_{ij}) \lVert_1),
\end{gather*}
here, we take token-wise linearization as an example, and $\mathrm{Lat}(\mathrm{Act_{ij}})$ denotes the latency of either GeLU or identity of $i$-th layer and $j$-th token, $T$ denotes the number of tokens in each layer. We can consider the balance between $\lambda$ and $\eta$ through the private inference latency:
$
    \frac{\lambda}{\eta} = \frac{\mathrm{Lat}(\mathrm{ATTN})}{\mathrm{Lat}(\mathrm{GeLU})}.
$

\textbf{Fuse two linear layers together.}
After GeLU linearization, we observe an opportunity for further latency reduction.
As shown in Figure \ref{fig:gelu_opt}, we fuse the linear layer before and after GeLU of a certain token, which reduces the communication associated with the MLP. 
Denote the input of a MLP block of $i$-th layer as $X^{(i)} \in \mathbb{R}^{B\times H\times W\times C}$, the linear layers as $\mathcal{W}^{(i)}_1\in \mathbb{R}^{C\times K}$ and $\mathcal{W}^{(i)}_2\in \mathbb{R}^{K\times C}$ , the output of MLP as $X^{(i+1)} \in \mathbb{R}^{B\times H\times W\times C}$. 
The input dimension of GeLU is $B\times H\times W\times K$, and $C, K$ are channel numbers of linear layers. 
Note that we generally set the ratio of MLP (i.e., $\frac{K}{C}$) to 2 or 4 in ViT models.
The linearized MLP can be formulated as
$
    X^{(i+1)} = (X^{(i)} \mathcal{W}^{(i)}_1) \mathcal{W}^{(i)}_2 = X^{(i)} (\mathcal{W}^{(i)}_1 \mathcal{W}^{(i)}_2).
$
In this way, $\mathcal{W}^{(i)}_1 \mathcal{W}^{(i)}_2$ is fused into $\mathcal{W}^{(i)}_f \in \mathbb{R}^{C \times C}$,
which optimizes two high-dimensional MatMuls to only a single low-dimensional MatMul, further reducing the latency of MLP.

\textbf{Adding additional activations.}
After GeLU removal and linear fusion, there is no non-linearity in these tokens. 
To boost the accuracy, we additionally add a non-linear function after the fused linear positions.
Here, we choose to add ReLU which incurs a cheaper cost than GeLU.

\section{Experiments}
\label{sec:exp}


\subsection{Experimental Setup}
\label{sec:setup}
\textbf{Experimental and MPC settings.}
In our experiments, we leverage the SecretFlow (SPU \cite{288747} V0.3.1b0)\footnote{\url{https://github.com/secretflow/secretflow}} framework for private inference, which is popular for privacy-preserving deep learning.
We adopt the SEMI-2K \cite{cramer2018spdz2k} protocol, which is a semi-honest two-party computation protocol\footnote{Although the actual latency of each operator may be different for different MPC protocols, our proposed algorithm can be generally applied.}.
We follow Cheetah \cite{huang2022cheetah} and use the WAN mode for communication.
Specifically, the communication bandwidth between the cloud instances is set to 44 MBps 
and the round-trip time is set to 40ms.
Our experiments are evaluated on an Intel Xeon CPU@2.40 GHz with 62 GB RAM.

\textbf{Model architectures and datasets.}
Many researches have studied training ViTs on small datasets~\cite{hassani2021escaping,gani2022train,huynh2022vision,liu2021efficient,lee2022improving}.
We consider two ViT architectures on three commonly used datasets following \cite{hassani2021escaping}. 
For the CIFAR-10/100 dataset, we set
the ViT depth, \# heads, hidden dimension, and patch size to 7, 4, 256, and 4, respectively.
For Tiny-ImageNet dataset, we set the ViT depth, \# heads, hidden dimension, and patch size to 9, 12, 192, and 4, respectively.

\textbf{Searching settings.} We run the NAS algorithm for 300 epochs with AdamW optimizer and a cosine learning rate across three datasets. We set $\epsilon = 10^{-8}$ to avoid the zero denominator in RSAttn, and $\lambda = 10^{-5}$. 

\textbf{Training settings.} 
Following~\cite{hassani2021escaping}, we train the searched heterogeneous ViTs for 600 epochs on CIFAR-10/100 and 300 epochs on Tiny-ImageNet. 
We leverage data augmentations as \cite{hassani2021escaping}.
For KD, we set the temperature $\mathcal{T}$ to 1, and set 
$\chi, \beta, \gamma$ to 1 as well.
To explore the trade-off between inference accuracy and latency,
we set $\mu$ to $\{0.1, 0.3, 0.5, 0.7\}$.

\subsection{Comparison with Prior-Art Efficient Attentions}

\begin{figure*}[!tbp]
    \centering
    \includegraphics[width=0.95\linewidth]{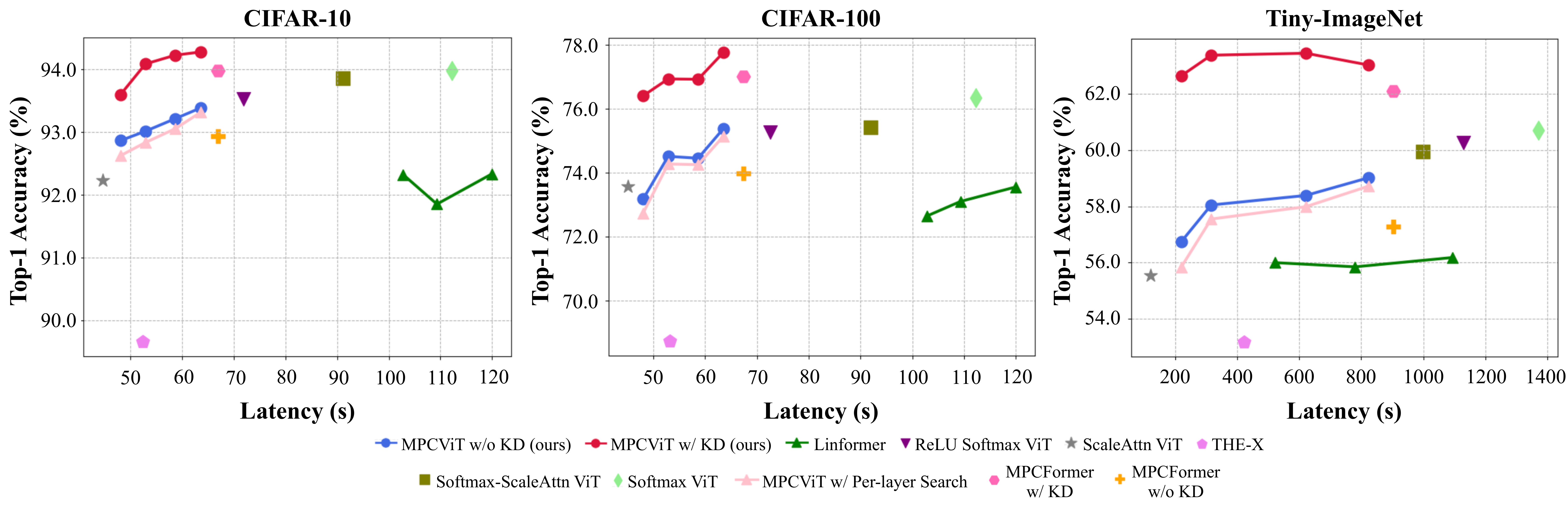}
    \caption{Top-1 Accuracy and inference latency comparison with baseline ViTs across three datasets.}
    \label{fig:main_result}
\end{figure*}

\textbf{Baselines.}
We compare MPCViT with prior-art models on the inference accuracy and
efficiency. 
We compare with MPCFormer \cite{li2022mpcformer}, Linformer \cite{wang2020linformer,wang2022characterization}, THE-X \cite{chen-etal-2022-x} and ViT variants in Table \ref{tab:attention_variants_operation}.
Linformer is studied in \cite{wang2022characterization} as an efficient
Transformer variant for MPC and the compression rate of attention dimension is also defined as $\mu$.
For networks compared with \method, we only modify attention layer and keep MLP unchanged as the baseline ViT.
For \method$^+$, we optimize attention and MLP simultaneously.

\textbf{Results and analysis.}
The main results are shown in Figure \ref{fig:main_result}, where we report top-1 accuracy across three datasets and the network inference latency.
The main findings are as follows:
\begin{itemize}
    \item[1)] \method~outperforms prior-art methods. Without KD, on Tiny-ImageNet, \method~with $\mu = 0.1$ outperforms Linformer
        with $\mu=0.7$ by 0.57\% better accuracy with $4.9 \times$ latency reduction;
        Compared with THE-X, \method~achieve 3.6\% accuracy improvement with 1.9$\times$ latency reduction;
    \item[2)] with proper KD, \method~even achieves 1.9\% and 1.3\% better accuracy with 6.2$\times$ and 2.9$\times$ latency reduction with $\mu=0.1$ and $0.3$, compared with baseline ViT and MPCFormer on
        Tiny-ImageNet, respectively;
    \item[3)] we mix Softmax attention and ScaleAttn.
        Although it has a slightly higher accuracy in plaintext inference,
        it incurs at least 1.4$\times$ and 1.2$\times$ latency on CIFAR-10/100 and Tiny-ImageNet compared with \method. 
\end{itemize}

\begin{figure}[!tbp]
    \centering
    \includegraphics[width=\linewidth]{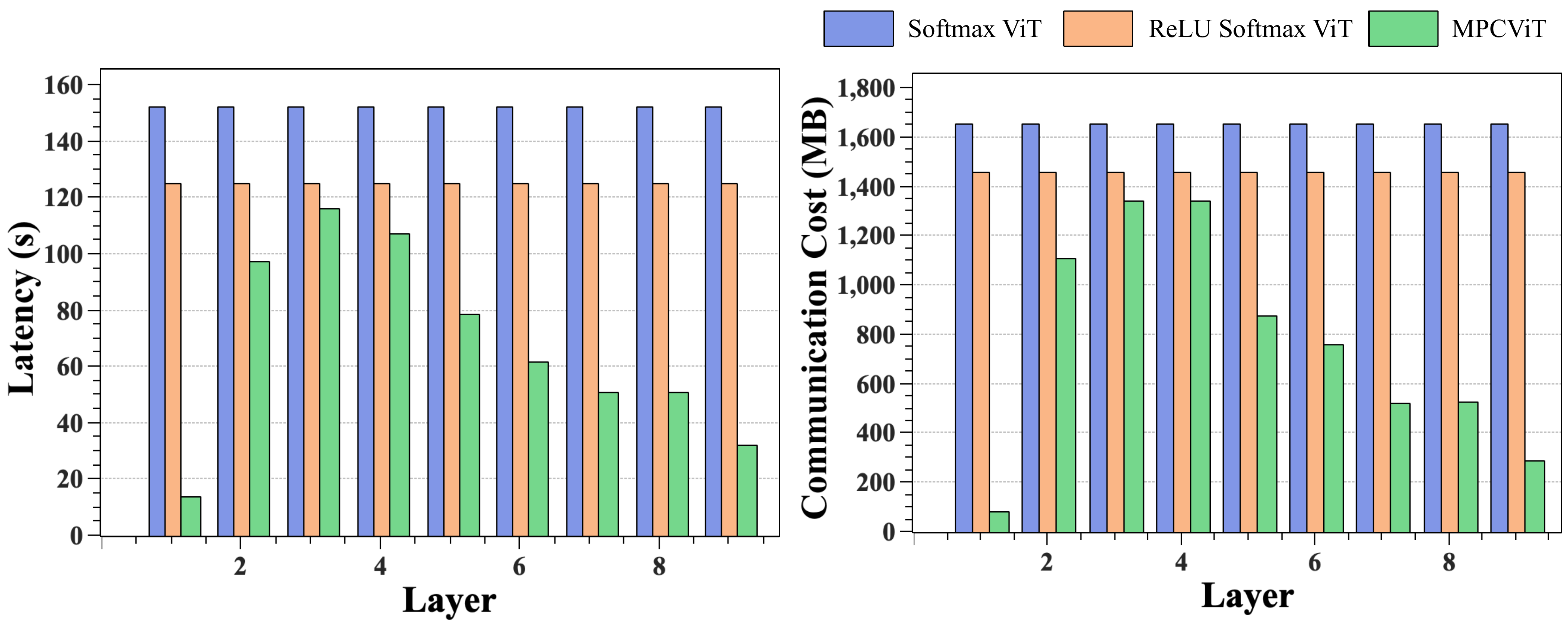}
    \caption{Latency and communication cost comparison of different layers in the 9-layer 12-head ViT with $\mu = 0.5$.}
    \label{fig:latency_communication}
    \vspace{-10pt}
\end{figure}

We visualize the latency and communication cost comparison of Softmax ViT, ReLU Softmax ViT, and \method~in Figure~\ref{fig:latency_communication}. We can observe the latency
changes proportionally with the communication. Also, we observe for \method,
RSAttn in middle layers (e.g., layer 2 to layer 5) tend to be preserved while
RSAttn in early and final layers are likely to be replaced with ScaleAttn.

\subsection{Comparison on Structure Granularities}

The choice of search space is important for \method, 
so we compare different structure granularities of search space proposed in~$\S$\ref{subsec:search_space}.
We use the CIFAR-10 dataset for the comparison.
As shown in Table~\ref{tab:granualarity}, the head-wise search space achieves the best accuracy.
We hypothesize although row-wise structure leads to the largest search space, such a fine-grained search space may introduce difficulties in NAS,
and hence, leads to network architecture with inferior accuracy. Head-wise search space achieves the right balance between flexibility and optimizability.

\begin{table}[htbp]
  \centering
  \caption{Top-1 accuracy (\%) comparison with different structure granularities of attention search space with different $\mu$.}
  \label{tab:granualarity}
  \scalebox{0.8}{
    \begin{tabular}{cccc}
    \toprule
    \multicolumn{1}{l}{Granularity} & \multicolumn{1}{l}{Layer-wise} & \multicolumn{1}{l}{Row-wise} & \multicolumn{1}{l}{Head-wise} \\
    \midrule
   $\mu=0.5$ &   93.01    &   93.16    & 93.21 \\
    $\mu=0.7$ &  93.32     &   93.13     &  93.38  \\
    \bottomrule
    \end{tabular}%
    }
\end{table}%

\subsection{Ablation Study of \method}
\label{ablation}

\textbf{Contribution of KD.}
One of our key techniques for training heterogeneous ViT is KD. 
We enumerate different combinations of KD and the results are shown in Table~\ref{tab:ablation_KD}.
We find that: 
1) both logits-based and feature-based KD significantly improve the performance of ViT compared to using cross-entropy loss only; and
2) combining the two KD losses further improves the accuracy, especially on a larger dataset.
The above findings indicate the importance of KD and the indispensability of three parts of
$\mathcal{L}_{train}$.

\begin{table}[!tb]
    \centering
    \caption{Top-1 Accuracy (\%) comparison of different combinations of training loss function with $\mu=0.7$.}
    \scalebox{0.65}{
        \begin{tabular}{ccccc}
        \toprule
            CE & Logits-based KD & Feature-based KD & CIFAR-10 & Tiny-ImageNet \\
        \hline
            \checkmark &  &  & 93.38 & 59.02\\
            \checkmark & \checkmark & & 94.18 & 62.39 \\
            \checkmark & & \checkmark & 94.12 & 62.26   \\
            & \checkmark & \checkmark & 94.14  & 61.80  \\
            \hdashline
            \checkmark & \checkmark & \checkmark & \textbf{94.27} & \textbf{63.03} \\
            \bottomrule
        \end{tabular}
    }
    \label{tab:ablation_KD}
\end{table}

\begin{figure}[!tbp]
    \centering
    \includegraphics[width=\linewidth]{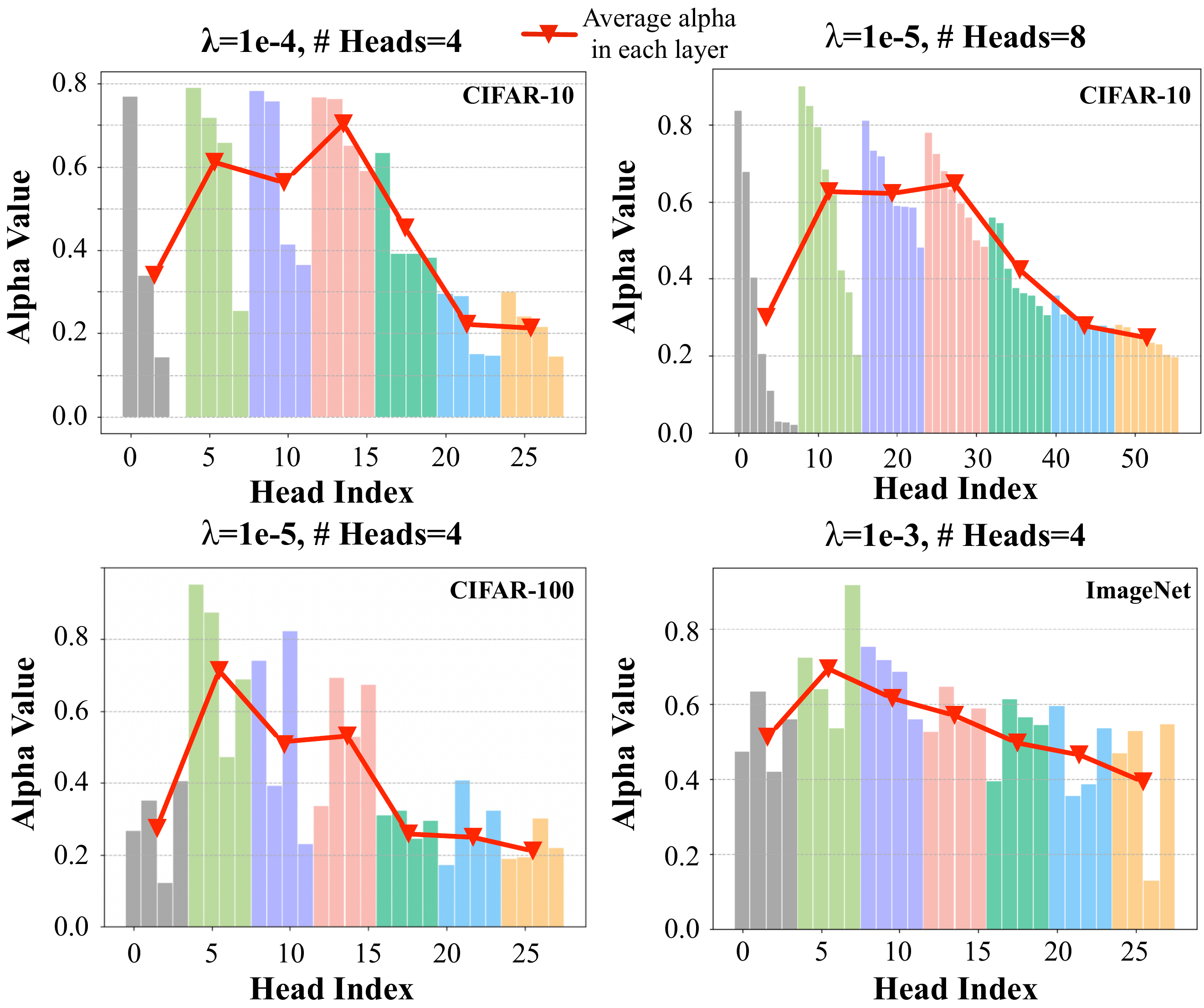}
    \caption{The distribution of architecture parameter $\alpha$ for each head under different $\lambda$ and different \# heads on different datasets. We draw the average $\alpha$ in each layer to show the similar trend.}
    \label{fig:alpha_distribution}
\end{figure}

\textbf{Consistency and scalability of NAS algorithm.}
We hope our NAS algorithm to be robust against hyper-parameter choices and datasets.
To analyze this consistency of our algorithm, we adjust the coefficient $\lambda$ to three different values, i.e., $10^{-3}, 10^{-4}, 10^{-5}$, and train a 7-layer ViT with 4 heads. 
We visualize the distribution of the architecture parameter $\alpha$ in Figure \ref{fig:alpha_distribution}, and we find that $\alpha$ in each layer has a quite similar trend under different $\lambda$'s.
Furthermore, when we change the number of heads to 8, or even train the same architecture on different datasets like CIFAR-100 and ImageNet,
the distribution still shows a similar trend, empirically proving the consistency and scalability of our proposed NAS algorithm.
Appendix \ref{supp:alpha_supp} shows more distribution trends under other settings.

\textbf{Comparison with per-layer search.}
We also compare \method~with per-layer search which replaces RSAttn heads in each layer uniformly given a certain $\mu$ across three datasets. 
As shown in Figure~\ref{fig:main_result}, \method~achieves a better Pareto front for accuracy and efficiency compared to the per-layer NAS.
The results demonstrate the importance to search and replace RSAttn heads across different layers.

\subsection{\method$^+$ Evaluation}
\label{exp:mpcvit+}

\begin{figure}[!tbp]
    \centering
    \includegraphics[width=\linewidth]{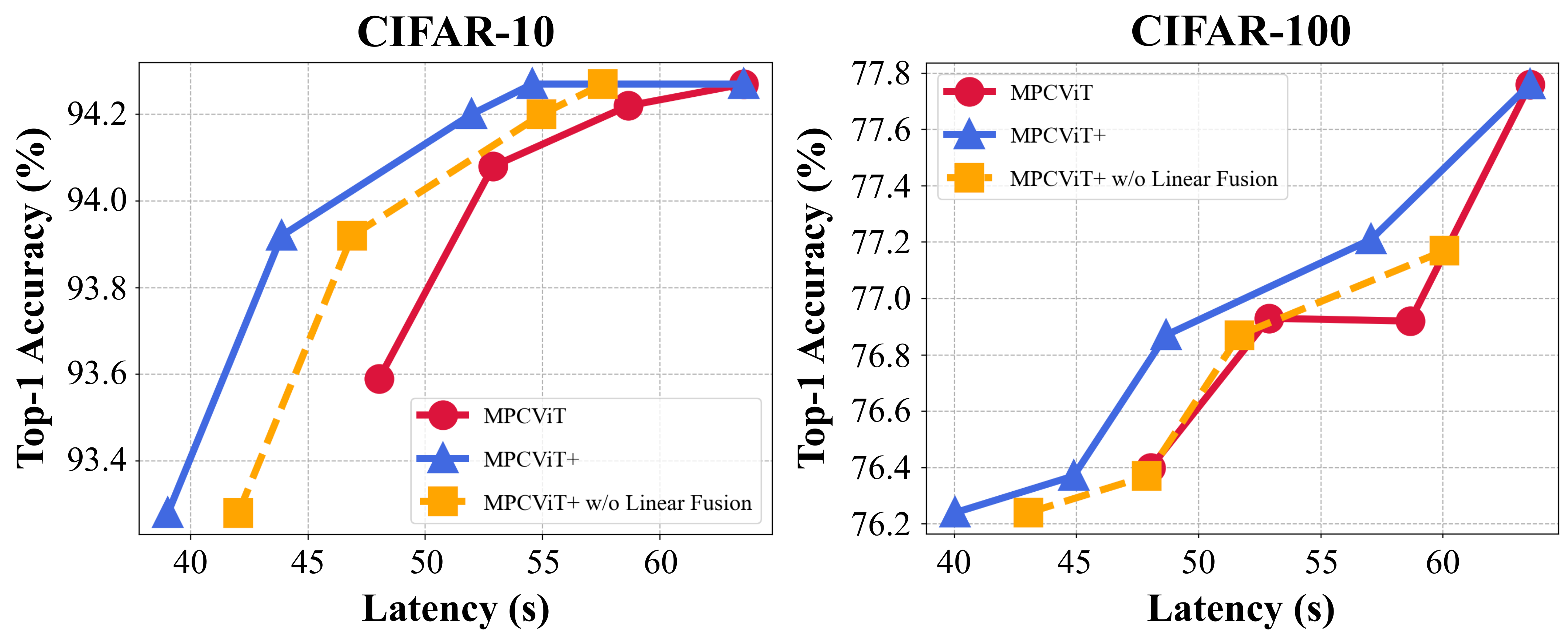}
    \caption{Comparison between \method~and \method$^+$.}
    \label{fig:mpcvit+}
\end{figure}

As shown in Figure \ref{fig:main_result}, when $\mu$ decreases to 0.1, the major latency bottleneck comes from GeLU in MLP.
We evaluate \method$^+$ for GeLU reduction on CIFAR-10/100 datasets with KD, which correspond to the ViT models with significant GeLU latency.
We take token-wise optimization as an example (exploration for other granularity, i.e., layer-wise is shown in Appendix \ref{supp:mpcvit+_layer}), and control the GeLU linearization rate using a threshold $\sigma$ such that $\beta \leftarrow 0 (\beta \leq \sigma)$ while $\beta \leftarrow 1 (\beta > \sigma)$.
As shown in Figure \ref{fig:mpcvit+}, GeLU linearization leads to a little accuracy degradation, indicating that GeLU plays an important role in ViTs.
\method$^+$ further reduces the latency on the basis of attention optimization, which proves the effectiveness of \method$^+$.

\textbf{Ablation study of additional ReLU.}
To analyze the help of adding additional ReLU after the fused linear layer, we fix $\sigma=0.75$ and perform the ablation study for \method$^+$ under different latency constraints on CIFAR-10 dataset in Table \ref{tab:abl_mpcvit+}.
The results show that additional ReLU improves the representation ability of \method$^+$.

\begin{table}[htbp]
  \centering
  \caption{Ablation study of adding additional ReLU after the fused linear layer under different latency constraints.}
  \label{tab:abl_mpcvit+}
  \scalebox{0.7}{
    \begin{tabular}{ccccc}
    \toprule
    \multirow{2}[4]{*}{$\mu$} & \multicolumn{2}{c}{0.3} & \multicolumn{2}{c}{0.7} \\
\cmidrule(r){2-3} \cmidrule(r){4-5}  & Accuracy (\%) & Latency (s) & Accuracy (\%) & Latency (s) \\
    \midrule
    \method$^+$ &   93.92    &  43.88        &   94.27    & 54.56  \\
    w/o ReLU &   93.72    &  41.37        &   93.81    & 52.11 \\
    \bottomrule
    \end{tabular}%
    }
\end{table}%

\section{Conclusion}
\label{sec:conclu}

In this paper, we present an MPC-friendly ViT family, dubbed \method,
to enable accurate yet efficient ViT inference in MPC.
We propose a heterogeneous attention optimization space, and design an MPC-aware differentiable NAS algorithm to explore the search space.
To further boost the inference efficiency, we propose \method$^+$ to jointly optimize attention and MLP.
Extensive experiments 
demonstrate that \method~consistently
outperforms prior-art ViT architectures with both lower latency and higher accuracy.
Our method can be further applied to large language models (LLMs) including BERT and GPT.

\section*{Acknowledgement}

This work is supported in part by the NSFC (62125401) and the 111 Project (B18001).

{
\small
\bibliographystyle{ieee_fullname}
\bibliography{egbib}
}

\clearpage
\appendix
\section{Private Vision Transformer Inference}
\label{supp:priv_vit}

\subsection{Overview of Private ViT Inference}
\label{supp:overview}
We illustrate the framework of private ViT inference to help readers to better understand our architecture.
As shown in Figure \ref{fig:private_transformer}, the server holds the model weight while the client holds the input data.
The client only send the secret share of data to the server to keep data private and the server keeps the weight private. 
In MPC, two parties, i.e., server and client, compute functionalities, e.g., linear and non-linear layers jointly with MPC protocols like Cheetah \cite{huang2022cheetah} and SEMI-2K \cite{cramer2018spdz2k}, and finally the client learns the inference results without extra information about the model weights while the server does not learn any information about the client's data.

\subsection{ViT Architecture}
\label{supp:vit_arch}
For the ViT~\cite{dosovitskiy2020image} architecture, it takes image patches as input and is composed of an input projection layer, a stack of Transformer layers, and a task-specific multi-layer perceptron (MLP) head. 
Each layer consists of an multi-head attention (MHA) layer and an MLP block. 
With MPC protocols, we have demonstrated that the communication bottleneck mainly comes from non-linear Softmax and GeLU.

\begin{figure}[!tbp]
    \centering
    \includegraphics[width=0.85\linewidth]{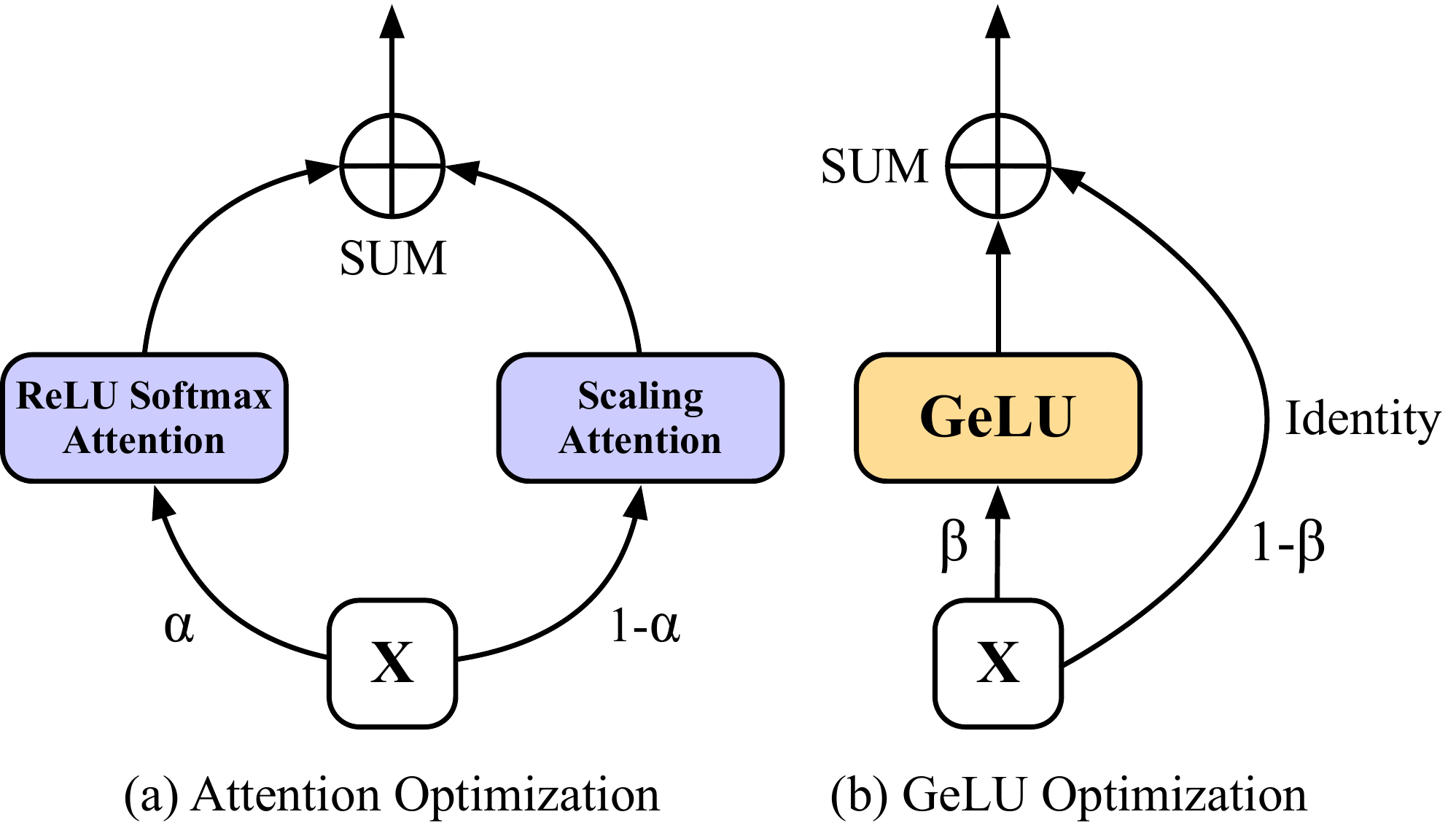}
    \caption{Visualization of our proposed \method~and \method$^+$.} 
    \label{fig:optimization}
\end{figure}

\paragraph{Softmax}
Softmax includes max, exponential and reciprocal operations, all of which are very expensive in MPC:
\begin{equation*}
    \mathrm{Softmax}(x_i) = \frac{e^{x_i - x_{max}}}{\sum_{j=1}^n e^{x_j - x_{max}}}.
\end{equation*}
Note that max is widely used in Softmax to improve numerical stability~\cite{wang2022characterization}.
Figure \ref{fig:optimization}(a) shows our method to optimize Softmax in MHA.

\paragraph{GeLU}
GeLU is an activation function based on the Gaussian error function \cite{hendrycks2016gaussian}, which is defined as:
\begin{equation*}
    \mathrm{GeLU}(x) = x \cdot \frac{1}{2} [1 + \mathrm{erf}(\frac{x}{\sqrt{2}})],
\end{equation*}
where $\mathrm{erf}(\cdot)$ is the Gaussian error function. For MPC, GeLU is usually approximated with tanh:
\begin{equation*}
    \mathrm{GeLU}(x) = 0.5x(1 + \mathrm{tanh}[\sqrt{\frac{2}{\pi}}(x + 0.044715x^3)]),
\end{equation*}
or
\begin{equation*}
    \mathrm{GeLU}(x) = x\cdot\sigma(1.702x).
\end{equation*}

Figure \ref{fig:optimization}(b) shows our method to optimize GeLU in MLP.

\paragraph{Exponential}
Exponential is used in Softmax, but exponential cannot be directly computed in MPC, so it is generally approximated as 
\begin{equation*}
    e^x = \lim_{n\rightarrow \infty}(1 + \frac{x}{2^n})^{2^n},
\end{equation*}
where $n$ is the number of approximation iterations.

\paragraph{Reciprocal}
Reciprocal is widely used in various functions, including Softmax, ReLU Sofmax, 2Quad, etc.
Reciprocal in MPC is usually approximated using Newton-Raphson iteration \cite{akram2015newton}:
\begin{equation*}
    \frac{1}{x} = \lim_{n\rightarrow \infty}y_n = y_{n-1}(2-xy_{n-1}),
\end{equation*}
where $y_0(x)=3e^{0.5-x}+0.003$ which makes the approximation work for a large input domain \cite{knott2021crypten}.

\begin{figure}[!tbp]
    \centering
    \includegraphics[width=\linewidth]{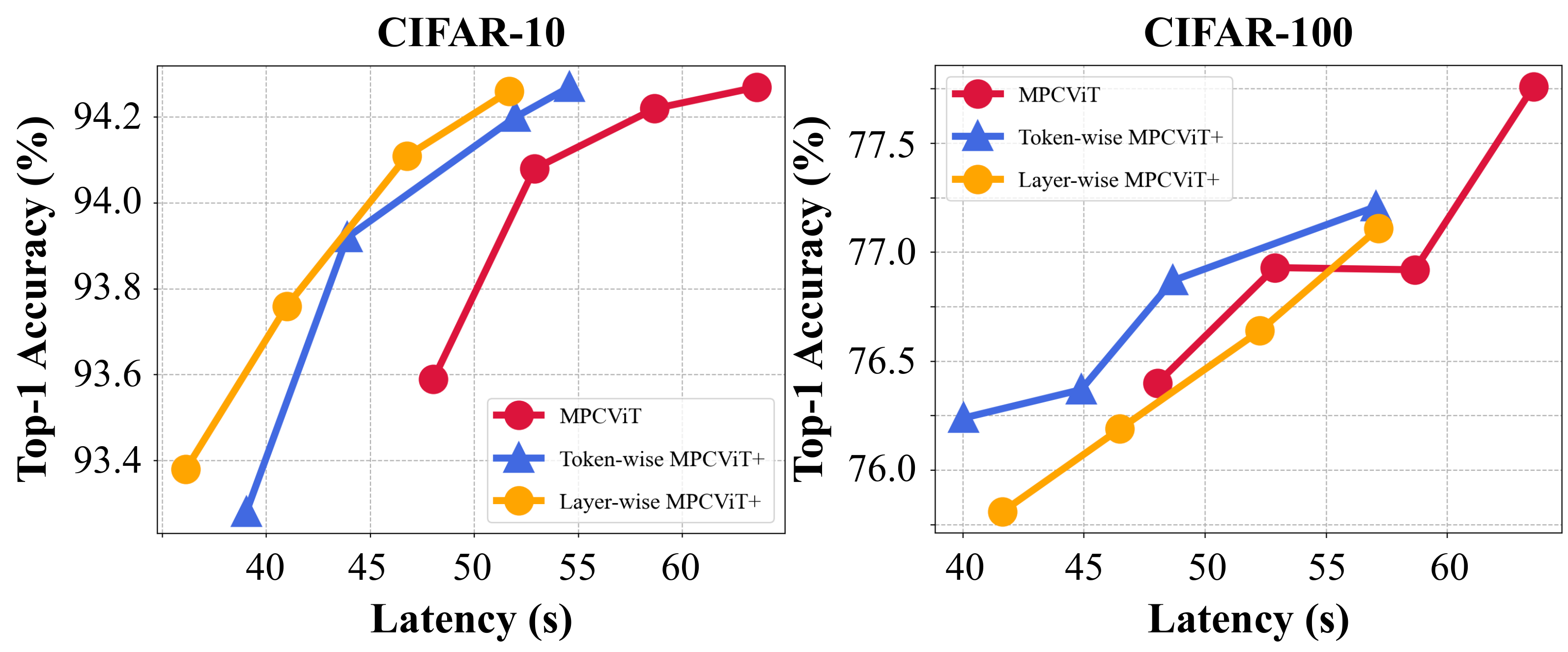}
    \caption{Comparison of layer-wise and token-wise GeLU optimization of \method$^+$.} 
    \label{fig:mpcvit+_layer}
\end{figure}

\section{Cryptographic Primitives for MPC}
\label{supp:crypto}
In this section, we briefly describe the relevant cryptographic primitives for MPC.
From the description, we can better understand the communication overhead brought from MPC during model inference.

\subsection{Additive Secret Sharing}
Additive secret sharing is widely used in arithmetic secret sharing (AS)~\cite{shamir1979share}.
Specifically, for an $l$-bit value $x \in \mathbb{Z}_{2^l}$, it is additively shared between two parties, denoted as $\langle x \rangle^l_0$ and $\langle x \rangle^l_1$, respectively, such that $x = \langle x \rangle^l_0 + \langle x \rangle^l_1 \mod 2^l$ (where $+$ denotes addition in $\mathbb{Z}_{2^l}$). 
AS is generally implemented by generating a pair $(r, x-r)$, where $r$ is a random number.
As illustrated on the left side of Figure \ref{fig:private_transformer}, we use the share algorithm $\mathrm{Share}^l(x)$ to split an input into two shares.
Conversely, we use reconstruction algorithm $\mathrm{Reconst}^l(\langle 
x \rangle^l_0, \langle 
x \rangle^l_1)$ recover the actual result to the client.
Note that, in AS, the communication of addition operation is free because addition can be locally computed.

\subsection{Oblivious Transfer}
Oblivious Transfer (OT) is the central cryptographic primitive for building MPC protocols to realize secure ViT private inference.
OT~\cite{naor2005computationally,asharov2013more} enables the receiver to choose one message obliviously from a set of messages sent from the sender without revealing his choice. 
For 1-out-of-$k$ OT, 
the sender holds $k$ $l$-bit messages $m_0, m_1, ... , m_{k-1} \in \{0, 1\}$ and the receiver holds a choice bit $b \in [k]$. 
At the end of OT protocol, the receiver learns $m_b$ but cannot learn any other massages, while the sender learns nothing.
Correlated OT (COT) is another form of OT, and 1-out-of-2 correlated OT
is widely used, e.g., SiRNN~\cite{rathee2021sirnn}.
Specifically, the sender inputs a correlation $x \in \mathbb{Z}_{2^l}$ and the receiver inputs a choice bit $b \in \{0, 1\}$. The protocol generates a random value $r \in \mathbb{Z}_{2^l}$ to the sender and $-r + b \cdot x$ to the receiver.
$\binom{k}{1}$-COT$_l$ requires $(2\lambda + kl)$-bit and 2 rounds communication.

\subsection{Garble Circuit}
Garble Circuit (GC)~\cite{yao1982protocols} enables two parties to jointly compute an arbitrary function $f(\cdot)$ without revealing their private information.
GC has three main phases: 1) garbling, 2) transferring and 3) evaluation.
First, the function $f(\cdot)$ is represented as a boolean circuit $C$. 
Then, the Garbler encoded the boolean circuit as a garbled circuit $\Tilde{C}$ and a set of
input-correspondent labels in the first phrase.
After garbling phrase, the Garbler sends $\Tilde{C}$ to another party who acts as the Evaluator together with
the correct labels for the input wires of the circuit.
The Evaluator computes the circuit gate-by-gate and produces an encoding of the output.
Finally, the Evaluator shares this encoding with the Garbler and learns the actual plaintext result.

\section{Details of Attention Variants}
\label{supp:detail_attn}

In this section , we formally describe the formulations of different attention variants mentioned in $\S$\ref{sec:motivating}.

\paragraph{Linformer \cite{wang2020linformer}}
\cite{wang2022characterization} takes Linformer as an efficient Transformer variant because Linformer significantly reduces the dimenstion of matrix $QK^T$.
\begin{equation*}
    \mathrm{Linformer}(Q, K, V) = \mathrm{Softmax}(\frac{Q(EK)^T}{\sqrt{d_k}}) \cdot (FV),
\end{equation*}
where $Q, K, V \in \mathbb{R}^{n \times d_k}$ are queries, keys and values, respectively. 
$E, F \in \mathbb{R}^{k\times n}$ are two linear projection matrices added on $K, V$ to compress the tensor size of $QK^T$.

\paragraph{ReLU/ReLU6 Attention}
ReLU/ReLU6 attention directly replaces Softmax with ReLU/ReLU6.
We take ReLU attention as an example and ReLU6 attention can be obtained by simply replacing ReLU with ReLU6 activation.
\begin{equation*}
    \mathrm{ReLUAttention}(Q, K, V) = \mathrm{ReLU}(\frac{QK^T}{\sqrt{d_k}}) \cdot V.
\end{equation*}

\paragraph{Sparsemax Attention}
\cite{martins2016softmax} proposes the Sparsemax activation function to enable to output sparse probabilities.
Sparsemax is defined as

\begin{align*}
    \begin{split}
        \mathrm{Sparsemax}(z)= \left \{
        \begin{array}{ll}
            1,                    & \mathrm{if} \quad t > 1; \\
            (t + 1)/2,     &  \mathrm{if} \quad -1 \leq t \leq 1;  \\
            0,                                 & \mathrm{if} \quad t < -1.
        \end{array}
        \right.
    \end{split}
\end{align*}

Thus, Sparsemax attention is defined as
\begin{equation*}
    \mathrm{SparsemaxAttention}(Q, K, V) = \mathrm{Sparsemax}(\frac{QK^T}{\sqrt{d_k}})\cdot V.
\end{equation*}

Note that Sparsemax can not only used for computing the output possibilities, but also for attention through replacing Softmax with Sparsemax in order to remove the expensive exponential.
However, Sparsemax requires more comparison operations.


\paragraph{XNorm Attention \cite{song2021ufo}}
XNorm is proposed by UFO-ViT \cite{song2021ufo} and is also called cross-normalization that normalizes $Q$ and $K^TV$ along two different dimensions to construct the linear attention:
\begin{align*}
    &\mathrm{XNormAttention}(Q, K, V) \\ =  &\mathrm{XNorm_{dim=filter}}(Q) (\mathrm{XN_{dim=space}}(K^TV)),
\end{align*}
\begin{align*}
    \mathrm{XN}(a) := \frac{\gamma a}{\sqrt{\sum_{i=0}^h \lVert a \Vert^2}},
\end{align*}
where $\gamma$ is a learnable parameter and $h$ is the hidden dimension.

\paragraph{2Quad Attention \cite{li2022mpcformer}}
2Quad approximation is proposed by MPCFormer \cite{li2022mpcformer}, which replaces $e^x$ with $(x+c)^2$ as follows:
\begin{equation*}
    \mathrm{2QuadAttention}(Q, K, V) = \frac{(\frac{QK^T}{\sqrt{d_k}} + c)^2}{\sum_{i=1}^n(\frac{QK^T}{\sqrt{d_k}} + c)^2_i} \cdot V.
\end{equation*}
In our experiments, we set $c$ to a very small value to make the training process robust.

\section{The Algorithm Flow of \method}
\label{supp:algo}
Here, we show the algorithm details of our proposed \method~pipeline.
 As shown in Algorithm \ref{alg:pipeline}, the pipeline is mainly divided into two steps: 
 search and retrain.
 We first initialize a ReLU Softmax ViT and jointly optimize the network $\theta$ and architecture parameter $\alpha$.
 After searching, we selectively replace a set of ReLU Softmax attention with an MPC-eficient Scaling attention based on the \textit{top-k} rule.
 Then, in order to boost the performance of \method~with heterogeneous attention, we retrain the ViT with knowledge distillation.
 The algorithm flow is almost the same with \method$^+$, and we can jointly optimize network $\theta$ and two architecture parameters $\alpha, \beta$ during the search.

 \begin{table}[!tbp]
  \centering
  \caption{Comparison of \method~and ReLU Softmax ViT with different number of heads via head pruning method.}
  \label{tab:heads}
  \scalebox{0.75}{
    \begin{tabular}{ccccc}
    \toprule
    \multicolumn{1}{c}{\multirow{2}[4]{*}{Dataset}} & \multicolumn{2}{c}{CIFAR-10} & \multicolumn{2}{c}{CIFAR-100} \\
\cmidrule(r){2-3}\cmidrule{4-5}          & \multicolumn{1}{l}{Accuracy (\%)} & \multicolumn{1}{l}{Latency (s)} & \multicolumn{1}{l}{Accuracy (\%)} & \multicolumn{1}{l}{Latency (s)} \\
    \midrule
    1-head &   92.48    &  50.88     &  73.25     &  51.12 \\
    2-head &  92.83     &  57.65     &  73.99     &  57.84  \\
    3-head &  93.03     &  66.21     &  74.61     &  66.70  \\
    \method &   \textbf{93.38}    &   63.56    &  \textbf{75.38}     &  63.79  \\
    \bottomrule
    \end{tabular}%
    }
\end{table}

\begin{algorithm*}[!tbp]
	\caption{Pipeline of Our Proposed \method}
	\label{alg:pipeline}
	\KwIn{ViT with ReLU Softmax attention: $f_{\theta}$; ratio of RSAttn budget $\mu$; searching epochs: $E_s$; training epochs: $E_t$; Lasso coefficient: $\lambda$; total number of ViT heads: $N$.}

	Initialize the architecture parameter $\alpha = 1.0$ for all attention heads in $f_{\theta}$.
	
	$\overline{\theta} \leftarrow (\theta, \alpha)$.
	
	\While{epoch $\leq E_s$} {
        Compute loss: $\mathcal{L}_{search}$ with the $\ell$1-penalty term;
        
        Update $\overline{\theta}$ with AdamW optimizer;
        
        Adjust learning rate with the cosine scheduler.
	}
	
	Sort the alpha values across the heads in $f_{\overline{\theta}}$, and find the $\mu N$-th largest $\alpha$, denoted as $\alpha^*$.
	
	\If{$\alpha \geq \alpha^*$} {
	    $\alpha \leftarrow 1.0$
	}
	\ElseIf{$\alpha < \alpha^*$}{
	    $\alpha \leftarrow 0.0$
	}
	
	Obtain the searched heterogeneous ViTs under different latency constraints with binarized $\alpha: f_{{\theta}^\prime}$. 
	
	Fix $\alpha$ and retrain $f_{{\theta}^\prime}$ to improve its accuracy as follows:
	
	\While{epoch $\leq$ $E_t$} {
        Compute loss: $\mathcal{L}_{train}$ with two types of KD techniques, i.e., $\mathcal{L}_{logists}$ and $\mathcal{L}_{feature}$;
    
        Update ${\theta}^\prime$ with AdamW optimizer;
        
        Adjust learning rate with the cosine scheduler.
	}	

        \KwOut{Accurate and efficient MPC-friendly ViTs with heterogeneous attention $f_{{\theta}^\prime}$.}

	Linearize ScaleAttn by scaling factor decomposition during inference time to accelerate computation. // Optional
\end{algorithm*}

\section{Layer-Wise VS. Token-Wise GeLU Optimizatin of \method$^+$}
\label{supp:mpcvit+_layer}

The choice of GeLU optimization can be different granualirties including layer-wise and token-wise, both of which support to fuse two linear layers for a better efficiency.
Experiments in $\S$\ref{exp:mpcvit+} use token-wise granularity as an example, and here we compare layer-wise and token-wise GeLU optimization for a better choice.
Since the proportion of GeLU and MatMul are small in the ViT model on Tiny-ImageNet, we here consider the model on CIFAR-10 and CIFAR-100 as shown in Figure \ref{fig:mpcvit+_layer}.
The results are evaluated with KD.
On CIFAR-10, \method$^+$ with layer-wise GeLU optimization has a little better Pareto front than token-wise optimization, while on CIFAR-100, token-wise optimization is a little better than layer-wise optimization.

\section{More Distributions of Attention Architecture Parameters}
\label{supp:alpha_supp}
\begin{figure}[!tbp]
    \centering
    \includegraphics[width=\linewidth]{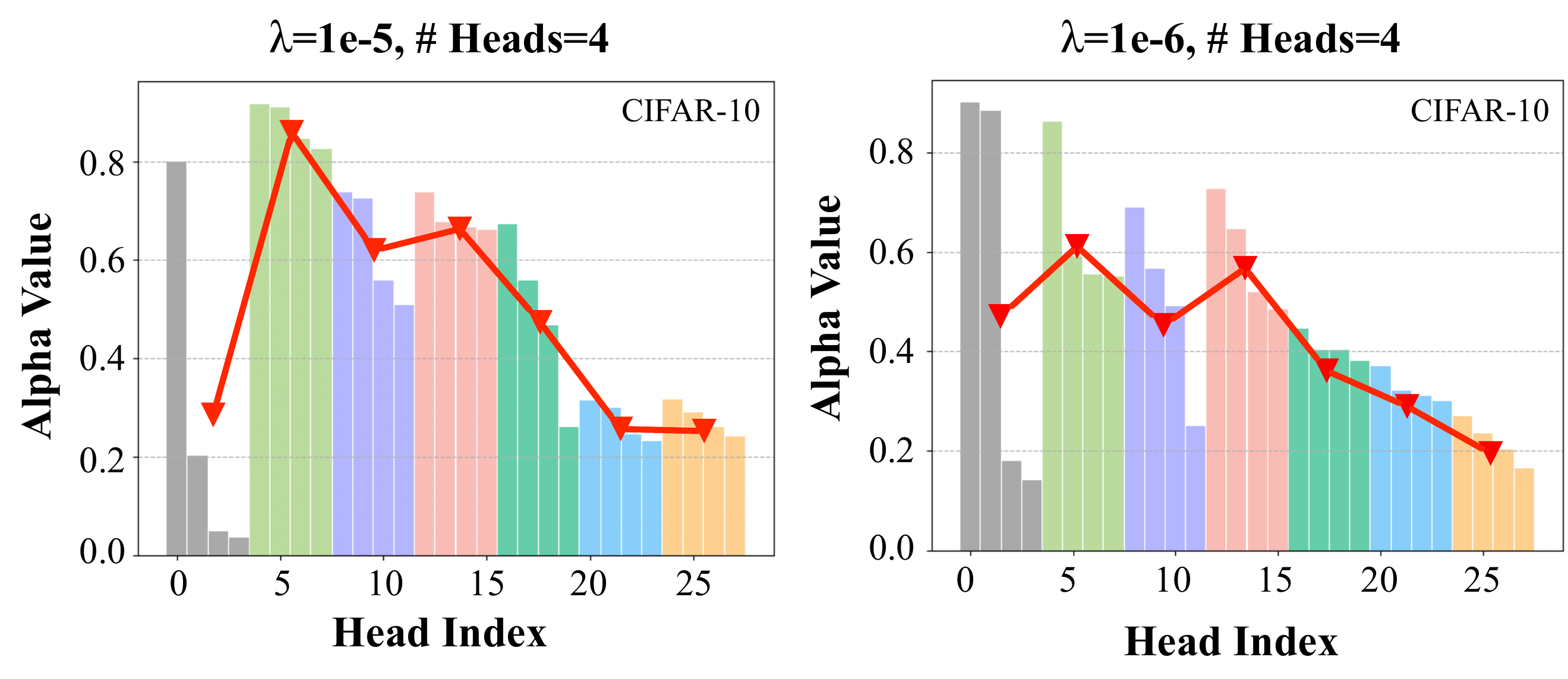}
    \caption{The distribution of architecture parameter $\alpha$ on CIFAR-10 with different Lasso coefficient $\lambda$.} 
    \label{fig:alpha_supp}
\end{figure}

In $\S$\ref{ablation}, we enumerate four situations to show the consistency and scalability of our NAS algorithm.
To empirically verify the consistency more sufficiently,
we supplement more cases of architecture parameter $\alpha$ for each attention head. 
On CIFAR-10, we fix the number of heads to 4 and modify the hyper-parameter $\lambda$ to even smaller values, i.e., $10^{-5}$ and $10^{-6}$.
As shown in Figure \ref{fig:alpha_supp}, the trend is still similar under different settings as Figure \ref{fig:alpha_distribution}.

\section{More Analysis: \method~Optimization VS. Pruning Method}
Our method is similar to pruning as we aim at removing ``unimportant'' modules in NNs.
Like \cite{cho2022selective}, here we analyze the advantage of our method.

Many methods \cite{xia2022structured,michel2019sixteen,voita2019analyzing} prune a subset of attention heads to improve the efficiency of Transformers.
The formulation of head pruning is defined as
\begin{equation*}
    \mathrm{MHA} = \sum_{i=1}^N z^{(i)} \mathrm{Att}(Q^{(i)}, K^{(i)}, V^{(i)}),
\end{equation*}
where $z^{(i)} \in \{0, 1\}$ is a mask variable for MHA.
However, this way losses the benefit of multi-head, leading to a worse representation ability.
Here, we give an example shown in Table \ref{tab:heads}.
Note that the ViT architecture on CIFAR-10/100 cannot support three heads since the hidden dimension is 256, so we just modify 256 to 258 with a negligible latency change.
As we can observe, \method~outperforms head pruning with higher accuracy and lower latency.
The result also indicates the necessity of including the ScaleAttn in MPCViT.
Instead of cutting attention heads, our method selectively replaces expensive attention with MPC-efficient attention without compromising the accuracy.

For \method$^+$, according \cite{cho2022selective}, our proposed GeLU linearization actually reduces the GeLU count while unstructured pruning still remains more GeLUs.
Compared with structured pruning, \method~maintains more parameters in the network \cite{cho2022selective}, achieving a better performance.

\end{document}